\documentclass[fullpage,11pt,palatino]{article}

\usepackage{amsmath,graphicx}
\usepackage{tabularx}
\usepackage{amsfonts}
\usepackage{stackrel}
\usepackage[ruled, linesnumbered]{algorithm2e}
\usepackage{amsthm}
\usepackage{bbm}
\usepackage{bm}
\usepackage[subpreambles=true]{standalone}
\usepackage{subcaption}

\usepackage{palatino}

\usepackage[usenames,dvipsnames]{xcolor}

\usepackage{tcolorbox}
\tcbuselibrary{skins,breakable}
\tcbset{enhanced jigsaw}

\definecolor{DarkRed}{rgb}{0.5,0.1,0.1}
\definecolor{DarkBlue}{rgb}{0.1,0.1,0.5}

\colorlet{YellowOrange}{RawSienna}

\usepackage{hyperref}%[backref=page]
\hypersetup{
colorlinks=true,
pdfnewwindow=true,
citecolor=ForestGreen,
linkcolor=DarkRed,
filecolor=DarkRed,
urlcolor=DarkBlue
}

\newtheorem{theorem}{Theorem}
\newtheorem{claim}{Claim}
\newtheorem{lemma}{Lemma}
\newtheorem{definition}{Definition}
\newtheorem{corollary}[theorem]{Corollary}
\newtheorem{corollaryforclaim}[claim]{Corollary}

\newtheorem{example}{Example}

\newtheorem*{lemma*}{Lemma}

\newcommand{\ind}{\textup{ind}}

\newcommand{\toShrink}{-.20cm}
\newcommand{\toShrinkEnu}{-.2cm}

\newcommand{\eps}{\ensuremath{\varepsilon}}

\newcommand{\bracket}[1]{\left[#1\right]}
\newcommand{\paren}[1]{\ensuremath{\left(#1\right)}\xspace}
\newcommand{\card}[1]{\left\vert{#1}\right\vert}

\newcommand{\floor}[1]{{\left\lfloor{#1}\right\rfloor}}
\newcommand{\prob}[1]{\Pr\paren{#1}}
\newcommand{\expect}[1]{\Exp\bracket{#1}}

\DeclareMathOperator*{\Exp}{\ensuremath{{\mathbb{E}}}}
\DeclareMathOperator*{\Prob}{\ensuremath{\textnormal{Pr}}}
\renewcommand{\Pr}{\Prob}

\newenvironment{tbox}{\begin{tcolorbox}[
		enlarge top by=5pt,
		enlarge bottom by=5pt,
		 breakable,
		 boxsep=0pt,
                  left=4pt,
                  right=4pt,
                  top=10pt,
                  boxrule=1pt,toprule=1pt,
                  colback=white,
                  arc=-1pt,
                  ]%%
	}
{\end{tcolorbox}}

\title{Near-linear Size Hypergraph Cut Sparsifiers}
\date{}

\author{Yu Chen\thanks{University of Pennsylvania, Philadelphia, PA, USA. Email: chenyu2@cis.upenn.edu.} \and Sanjeev Khanna\thanks{University of Pennsylvania, Philadelphia, PA, USA. Email: sanjeev@cis.upenn.edu.} \and Ansh Nagda\thanks{University of Washington, Seattle, WA, USA. Email: ansh@cs.washington.edu.}}

\begin{document}
\maketitle

\begin{abstract}
Cuts in graphs are a fundamental object of study, and play a central role in the study of graph algorithms. The problem of sparsifying a graph while approximately preserving its cut structure has been extensively studied and has many applications.
In a seminal work, Benczúr and Karger (1996) showed that given any $n$-vertex undirected weighted graph $G$ and a parameter $\eps \in (0,1)$, there is a near-linear time algorithm that outputs a weighted subgraph $G'$ of $G$ of size $\tilde{O}(n/\eps^2)$ such that the weight of every cut in $G$ is preserved to within a $(1 \pm \eps)$-factor in $G'$. The graph $G'$ is referred to as a {\em $(1 \pm \eps)$-approximate cut sparsifier} of $G$.

A natural question is if such cut-preserving sparsifiers also exist for hypergraphs.  Kogan and Krauthgamer (2015) initiated a study of this question and showed that given any weighted hypergraph $H$ where the cardinality of each hyperedge is bounded by $r$, there is a polynomial-time algorithm to find a $(1 \pm \eps)$-approximate cut sparsifier of $H$ of size $\tilde{O}(\frac{nr}{\eps^2})$. Since $r$ can be as large as $n$, in general, this gives a hypergraph cut sparsifier of size $\tilde{O}(n^2/\eps^2)$, which is a factor $n$ larger than the Benczúr-Karger bound for graphs. It has been an open question whether or not Benczúr-Karger bound is achievable on hypergraphs. In this work, we resolve this question in the affirmative by giving a new polynomial-time algorithm for creating hypergraph sparsifiers of size $\tilde{O}(n/\eps^2)$.

\end{abstract}

\section{Introduction}

In many applications, the underlying graphs are too large to fit in the main memory, and one typically builds a compressed representation that preserves relevant properties of the graph. 
Cuts in graphs are a fundamental object of study, and play a central role in the study of graph algorithms. Consequently, the problem of {\em sparsifying} a graph while approximately preserving its cut structure has been extensively studied (see, for instance,~\cite{Karger:1993:GMR:313559.313605, benczur1996approximating,Karger99, SpielmanT04, AhnG09, AhnGM12b,GoelKP12, BatsonSS12, AhnGM13, LeeS17, KapralovLMMS17, BansalST19, KapralovMMMNST20}, and references therein).
A cut-preserving sparsifier not only reduces the space requirement for any computation, but it can also reduce the time complexity of solving many fundamental cut, flow, and matching problems as one can now run the algorithms on the sparsifier which may contain far fewer edges.
In a seminal work, Benczúr and Karger~\cite{benczur1996approximating} showed that given any $n$-vertex undirected weighted graph $G$ and a parameter $\eps \in (0,1)$, there is a near-linear time algorithm that outputs a weighted subgraph $G'$ of $G$ of size $\tilde{O}(n/\eps^2)$ such that the weight of every cut in $G$ is preserved to within a multiplicative $(1 \pm \eps)$-factor in $G'$. The graph $G'$ is referred to as the {\em $(1 \pm \eps)$-approximate cut sparsifier} of $G$.

In this work, we consider the problem of cut sparsification for hypergraphs. A hypergraph $H(V,E)$ consists of a vertex set $V$ and a set $E$ of hyperedges where each edge $e \in E$ is a subset of vertices. The {\em }rank of a hypergraph is the size of the largest edge in the hypergraph, that is, $\max_{e \in E}\card{e}$. Hypergraphs are a natural generalization of graphs and many applications require estimating cuts in hypergraphs (see, for instance, ~\cite{CatalyurekA99,CatalyurekBDBHR09,HuangLM09,YamaguchiOTI15}).
Note that unlike graphs, an $n$-vertex hypergraph may contain exponentially many (in $n$) hyperedges.
This strongly motivates the question if cut-preserving sparsifiers in the spirit of graph sparsifiers can also be created for hypergraphs as this would allow algorithmic applications to work with hypergraphs whose size is polynomially bounded in $n$.

Kogan and Krauthgamer~\cite{kogan2015sketching} initiated a study of this basic question and showed that given any weighted hypergraph $H$, there is an $O(mn^2)$ time algorithm to find a $(1 \pm \eps)$-approximate cut sparsifier of $H$ of size $\tilde{O}(\frac{nr}{\eps^2})$ where $r$ denotes the rank of the hypergraph. Similar to the case of graphs, the {\em size} of a hypergraph sparsifier refers to the number of edges in the sparsifier. Since $r$ can be as large as $n$, in general, this gives a hypergraph cut sparsifier of size $\tilde{O}(n^2/\eps^2)$, which is a factor of $n$ larger than the Benczúr-Karger bound for graphs. 
Chekuri and Xu~\cite{Chekuri018} designed a more efficient algorithm for building a hypergraph sparsifier. They gave a near-linear time algorithm in the total representation size (sum of the sizes of all hyperedges) to construct a hypergraph sparsifier of size $\tilde{O}(nr^2/\eps^2)$ in hypergraphs of rank $r$, thus speeding up the run-time obtained in the work of Kogan and Krauthgamer~\cite{kogan2015sketching} by at least a factor of $n$, but at the expense of an increased sparsifier size.
It has remained an open question if the Benczúr-Karger bound is also achievable on hypergraphs, that is, do there exist hypergraph sparsifiers with $\tilde{O}(n/\eps^2)$ edges? In this work, we resolve this question in the affirmative by giving a new polynomial-time algorithm for creating hypergraph sparsifiers of size $\tilde{O}(n/\eps^2)$.

\begin{theorem} \label{thm:main}
    Given a weighted hypergraph $H$, for any $0<\eps<1$, there exists a randomized algorithm that constructs a $(1 \pm \eps)$-approximate cut sparsifier of $H$ of size $O(\frac{n\log n}{\eps^2})$ in $\tilde{O}(mn + n^{10}/\eps^7)$ time with high probability; here $n$ denotes the number of vertices and $m$ denotes the number of edges in the hypergraph.
\end{theorem}

It is worth noting that the size bound obtained in Theorem~\ref{thm:main} is the best possible to within a logarithmic factor even when the input is an {\em unweighted} hypergraph that only contains edges of rank $\Omega(n)$. Consider the following ``sunflower graph'' with $2n$ vertices, say, $v_1, v_2, ..., v_{2n}$,  and $n$ hyperedges.
For any $1 \le i \le n$, the $i_{th}$ hyperedge $e_i$ contains vertex $v_i$ along with the vertices $v_{n+1},v_{n+2},\dots,v_{2n}$. For any $1 \le i \le n$, the size of the cut $(\{v_i\}, V \setminus \{v_i\})$ is $1$ as $e_i$ is the unique edge cut by this cut. So any sparsifier for this graph must include every hyperedge. This in particular means that the bound in 
Theorem~\ref{thm:main} is the best possible to within a logarithmic factor even when one measures the total {\em representation size} of a hypergraph cut sparsifier, and not just the number of edges.

We now briefly describe the high-level idea behind the proof of Theorem~\ref{thm:main}. 
In the work of Benczúr and Karger~\cite{benczur1996approximating}, a graph sparsifier is constructed by sampling the edges with probabilities according to their {\em strengths}, a notion that captures the importance of an edge. Informally speaking, any edge that is among a small number of edges crossing some cut will have a high strength while any edge that does not participate in any small cuts will have a low strength. Once edges are sampled in this manner, a second key element in showing that the (appropriately weighted) sampled graph approximately preserves {\em every cut} in the original graph, is to establish a cut counting bound which shows that there can not be too many cuts that are within a given factor of the minimum cut size in the graph. This allows use of a union bound over all cuts to show that every cut is well-approximated. Kogan and Krauthgamer~\cite{kogan2015sketching} extend this elegant approach to constructing hypergraph sparsifiers. Similar to~\cite{benczur1996approximating}, they construct a hypergraph sparsifier by sampling hyperedges according to their strengths. A key point of divergence occurs in the second element, namely, the cut counting bound. As it turns out, number of cuts that are within a given factor of the minimum cut size, can be exponentially larger in the setting of hypergraphs\footnote{As a simple example (derived from an example in~\cite{kogan2015sketching}), consider a $n$-vertex hypergraph that contains a single hyperedge of size $n$ with weight $1$, as well as a clique on the $n$ vertices such that each clique edge has weight $1/n^2$. It is easy to see that the weight of a minimum cut in this graph is $1 + (n-1)/n^2 \approx 1$. On the other hand, all possible $2^n-1$ non-trivial partitions of the $n$ vertices gives us a cut of size at most $3/2$. This is an exponential increase compared to the graph setting where it is known that the number of cuts that are at most twice as big as the minimum cut is bounded by $O(n^4)$~\cite{Karger:1993:GMR:313559.313605}. Note the $2^n-1$ cuts created above not only correspond to distinct vertex partitions, but also have a distinct set of edges crossing them. Interestingly, the maximum number of distinct minimum cuts is the same in both graphs and hypergraphs, see, for instance, the work of Ghaffari, Karger, and Panigrahi~\cite{GhaffariKP17}.\label{footnote:cut-counting-example}}.
To compensate for this increase in the number of cuts, their algorithm samples edges at roughly $r$ times higher rate, resulting in a sparsifier of $\tilde{O}(nr)$ for hypergraphs of rank $r$. This size bound is essentially best possible by a direct execution of the Benczúr-Karger framework.

Our proof of Theorem~\ref{thm:main} follows the high-level idea of creating a suitable probability distribution 
over hyperedges, and then sampling them in accordance with this distribution. However, we construct our hyperdge sampling distribution by analyzing the interaction among hyperdeges at a finer granularity. 
In particular, we start by constructing an auxiliary graph $G$ where for each hyperedge $e$ in $H$, we add a clique $F_e$ whose vertex set is the same as the vertex set of the hyperedge $e$. The probability of sampling a hyperedge $e$ in $H$ is now determined by the strengths of the edges in the clique $F_e$. 
However, for this ``sparsification-preserving coupling'' between the graphs $G$ and $H$ to work, we can not directly use the graph $G$ but instead need to create a {\em non-uniform} weight assignment to the edges in $G$ that roughly ensures that the edges in $F_e$ have similar strengths in $G$. In particular, for any hyperedge $e$, the edges in $F_e$ may get assigned weights that now range from $0$ to the weight of the hyperedge $e$. This weight assignment scheme, referred to as a {\em balanced assignment}, and an algorithm to compute it efficiently, are the key technical insights in our work.
We note that the strategy of building sparsifiers of a hypergraph by the auxiliary graph $G$ is also used in~\cite{BansalST19} where the authors use this strategy to construct spectral hypergraph sparsifier. Unlike our scheme, however, the work in~\cite{BansalST19} assigns uniform weights to the edges in $F_e$. 

We conclude our overview by summarizing the three main technical steps involved in obtaining Theorem~\ref{thm:main} by executing the high-level idea and described above. In the first step, we assign weights to the edges in $G$ so that the edges in each clique $F_e$ have similar strengths. In general, this task might be impossible, but we get around this by working with a weaker condition, namely, we only require that all edges in $F_e$ that receive a positive weight have similar strengths. We design an iterative algorithm to achieve this goal, and show that it converges in polynomial time. In the second step, we prove that the hypergraph sparsifier constructed by sampling each hyperedge $e$ according to the strengths of edges in $F_e$ is indeed a good sparsifier for our input hypergraph. The proof of the second step follows the framework in~\cite{benczur1996approximating} at a high-level but a key challenge is to couple together the performance of a sparsifier in $H$ with the performance of a sparsifier in $G$. Together these two steps give us a polynomial-time algorithm for constructing a hypergraph sparsifier of size $\tilde{O}(n/\eps^2)$. However, the running time of the resulting algorithm is quadratic in terms of $m$, the number of hyperedges. Since in a hypergraph, the number of edges $m$ can be exponentially larger than $n$, in the third step, we present a way to speed up the algorithm so that the run-time has only a linear dependence on $m$.

Finally, we note that Theorem~\ref{thm:main} also yields a $\tilde{O}(n^2/\eps^2)$ space streaming algorithm for building a hypergraph sparsifier in a single-pass over an insertion-only stream. This can be done using a black-box technique for transforming cut sparsification algorithms into streaming algorithms whose space requirement is only slightly more than the sparsifier size (see Section 2.2 of~\cite{McGregor14}):

\begin{lemma} [\cite{McGregor14}] \label{lem:streaming}
    Given an algorithm that finds a $(1\pm \eps)$-approximate cut sparsifier of a hypergraph of size at most $f(n, \eps)$ with high probability, there exists a single-pass insertion-only streaming algorithm to compute a $(1\pm\eps)$-approximate cut sparsifier of size $2\log(m/n) \cdot f(n,\frac{\eps}{2\log(m/n)})$ that stores at most $2\log^2(m/n)\cdot f(n,\frac{\eps}{2\log(m/n)})$ hyperedges at any given time with high probability.
\end{lemma}

\begin{corollary}
    For any $0<\eps<1$, there exists a randomized insertion only streaming algorithm that constructs $(1 \pm \eps)$-approximate cut sparsifier of $H$ of size $O(\frac{n\log n \log^3(m/n)}{\eps^2})$ with high probability and stores only $O(\frac{n\log n \log^4(m/n)}{\eps^2})$ hyperedges, and hence uses $O(\frac{n^2 \log n \log^4(m/n)}{\eps^2})$ space in the worst-case. 
\end{corollary}

The above result improves upon the $\tilde{O}(n^3/\eps^2)$ space streaming algorithm in~\cite{kogan2015sketching} for building hypergraph sparsifiers in insertion-only streams. We note here that for hypergraphs of {\em constant} rank, an $\tilde{O}(n/\eps^2)$ space streaming algorithm is known~\cite{GuhaMT15} in dynamic streams where both insertion and deletion of hyperedges is allowed.

\smallskip
\noindent
{\bf Related Work:}
Spielman and Teng~\cite{SpielmanT04} introduced a natural strengthening of the notion of cut sparsifiers in graphs, called a {\em spectral sparsifier}. A $(1 \pm \eps)$-approximate spectral sparsifier of a graph $G(V,E)$ is a weighted graph $G'(V,E')$ such that for every vector $x \in \mathbb{R}^{n}$, we have 
$$  ~~~ |x^T L_{G'} ~x - x^T L_G ~x | \leq \eps (x^T L_G ~x) ,$$
where $L_G$ and $L_{G'}$  denote the Laplacian matrices of $G$ and $G'$, respectively. To see that the notion of spectral spasrifier only strengthens the notion of a cut sparsifier, observe that the cut sparsification requirement for any cut $(S, \bar{S})$ is captured by the definition above when we choose $x$ to be the $0/1$-indicator vector of the set $S$.
Batson, Spielman, and Srivastava~\cite{BatsonSS12} gave a polynomial-time algorithm that for every graph $G$, gives a weighted graph $G'$ with $O(n/\eps^2)$ edges such that $G'$ is a $(1 \pm \eps)$-approximate spectral sparsifier of $G$. Subsequently, Lee and Sun~\cite{LeeS17} gave an $O(m/\eps^{O(1)})$ time algorithm to construct a spectral graph sparsifier with $O(n/\eps^2)$ edges.

Very recently, Bansal, Svensson, and Trevisan~\cite{BansalST19} explored both the standard multiplicative error notion as well as a weaker notion of graph and hypergraph sparsification whereby the size of each cut $(S, \bar{S})$, is approximated to within an additive error that is bounded by $\eps (d|S| + {\rm vol}(S))$ where $d$ is the average degree in the graph, and ${\rm vol}(S)$ denotes the sum of degrees of vertices in $S$. It is easy to see that in general, the additive error term allowed in the weaker notion can be $\Omega(m)$ times larger than the multiplicative error even in connected graphs. Bansal {\em et al.} designed a randomized polynomial time algorithm that gives {\em unweighted} hypergraph sparsifiers of size $O(\frac{n \log (r/\eps)}{\eps^2 r} )$ for the weaker notion defined above. For the multiplicative error notion, they give a polynomial-time algorithm that outputs a weighted spectral sparsifier with $O(\frac{r^3}{\eps^2} \cdot n \log n)$ hyperedges. This latter result is in contrast to the recent result of Soma and Yoshida~\cite{SomaY19} who gave spectral hypergraph sparsifiers with $O(\frac{n^3 \log n}{\eps^2})$ hyperedges. 

There has also been extensive work on designing space-efficient streaming algorithms for cut sparsifiers as well as spectral sparsifiers for graphs, starting with the work of Ahn and Guha~\cite{AhnG09} who gave the first $\tilde{O}(n/\eps^2)$ space single-pass streaming algorithm to build a $(1 \pm \eps)$-approximate cut sparsifier in insertion-only streams. 
Ahn, Guha, and McGregor~\cite{AhnGM12a} introduced a powerful linear-sketching primitive for graph connectivity that led to the construction of graph sparsifiers using $\tilde{O}(n/\eps^2)$ space in the more general setting of dynamic streams where a graph is revealed as a sequence of edge insertions and deletions~\cite{AhnGM12b,GoelKP12}.
Subsequently, similar results have also been obtained for spectral sparsifiers in dynamic graph streams~\cite{AhnGM13, KapralovLMMS17, KapralovMMMNST20}.

\iffalse
Ghaffari, Karger, and Panigrahi prove that the number of minimum cuts in a hypergraph is at most ${n \choose 2}$, which is the same as the bound for graphs. However, as shown by the example in footnote \ref{footnote:cut-counting-example}, the cut counting bound for graphs no longer holds for does not hold for hypergraph when $\alpha>1$, the authors show that, for any $\alpha \ge 1$, there is a set $O(n^{\alpha/\delta})$ of cuts such that for any $\alpha$ min-cut $C$ in the hypergraph, there is a cut $C'$ in the graph that $C$ contains $(1-\delta)$ fraction of hyperedges in $C'$. Note that the reverse may not be true, the size of $C$ might be much larger than $C'$. The result can be used to build a sparsifier of a hypergraph that only preserves the min-cut size.
\fi

\medskip
\noindent
{\bf Organization:} We set up our notation and state some useful background results in Section~\ref{sec:prelims}. We present a detailed technical overview of our hypergraph sparsifier construction in Section~\ref{sec:overview}. 
In Section~\ref{sec:balance}, we give a polynomial-time algorithm to construct a balanced weight assignment, and in Section~\ref{sec:sparsifier}, we show how a balanced weight assignment can be used to create a hypergraph sparsifier with $O(\frac{n\log n}{\epsilon^2})$ edges. Finally, in Section~\ref{sec:group}, we present a way to speed-up our algorithm so that the final algorithm has only a linear dependence on $m$, completing the proof of Theorem~\ref{thm:main}.

\section{Preliminaries}
\label{sec:prelims}

\subsection{Notation}

A \emph{hypergraph} is defined as a pair $(V, E)$ of vertices and edges, where each edge in $E$ is a subset of $V$. In this paper, we allow parallel edges (that is, $E$ is a multiset). To emphasize this, we often refer to a graph/hypergraph as a multigraph/multihypergraph. Given a weight function $w$ that assigns a nonnegative weight to each edge in $E$, the triple $(V, E, w)$ is a weighted hypergraph. Notice that an unweighted graph/hypergraph can be thought of as a weighted graph/hypergraph with all weights equal to $1$.

Throughout the paper, we use ``graph'' to refer to standard graphs with edges of size $2$, and ``hypergraph'' to refer to graphs where edge sizes are arbitrary. We generally use the symbol $G$ to refer to standard graphs, and $H$ to refer to hypergraphs. Additionally, we generally use $f$ to denote an edge in a standard graph, and $e$ to denote an edge of a hypergraph. We will assume throughout that we are dealing with a hypergraph with at least $n$ edges, since otherwise, we can simply output $H$ as its own sparsification. Finally, the phrase ``with high probability'' means with probability $1 - 1/poly(n)$ for some large polynomial in $n$.

Given any weight function $w:S\rightarrow \mathbb{R}_{\geq 0}$, we extend it to also be a function on subsets of $S$ so that $w(S') = \sum_{e\in S'}w(e)$ for $S'\subseteq S$. Given a weighted graph/hypergraph $G = (V, E, w)$ and a subset of vertices $V'\subseteq V$, we define $G[V']$ to be the weighted subgraph/subhypergraph of $G$ induced by the vertices in $V'$.

A cut $C = (S, \bar{S})$ of a vertex set $V$ is any disjoint partition of $V$ into two sets such that neither of the sets are empty. Given a graph/hypergraph $G = (V,E,w)$ and a cut $C = (S,\bar{S})$, we denote by $\delta_G(S)$ the set of the edges crossing the cut $C$ in $G$. By definition, $\card{\delta(S)}$ is the number of edges crossing $C$ and $w(\delta(S))$ is the weight/size of $C$. A {\em $(1 \pm \eps)$-approximate cut sparsifier} of $G$ is a graph/hypergraph $G' = (V,E',w')$ with $E' \subseteq E$ such that 

$$\forall S \subseteq V, ~~~\card{ w'(\delta_{G'}(S))- w(\delta_G(S)) } \leq \eps w(\delta_G(S)).$$	

The following concentration bound can be found in \cite{fung2019general}:
    
\begin{lemma}[Theorem $2.2$ in \cite{fung2019general}]\label{chernoff}
    Let $\{x_1,\ldots, x_k\}$ be a set of random variables, such that for $1 \le i \le k$, each $x_i$ independently takes value $1/p_i$ with probability $p_i$ and $0$ otherwise, for some $p_i \in [0,1]$. Then for all $N\geq k$ and $\eps\in (0,1]$, 
    $$
    \prob{\card{\sum_{i\in [k]}x_i - k}\geq \eps N} \leq 2e^{-0.38\eps^2\cdot \min_i p_i\cdot N}
    $$
\end{lemma}

\subsection{Edge Strengths and the Cut Counting Bound}
We review some concepts and results that can be found in previous works on cut sparsifiers in standard graphs, which also play important roles in our algorithm. 

\begin{definition}
    Given a weighted graph $G$, a \emph{$k$-strong component} of $G$ is a maximal induced subgraph of $G$ that has minimum cut at least $k$.
\end{definition} 
\begin{lemma}[\cite{benczur1996approximating}]\label{lemma-laminar}
    Given a weighted graph $G=(V, F, w)$ and some real number $k$, the $k$-strong components of $G$ partition $V$. Given another real number $k'\geq k$, the $k'$-strong components of $G$ are a refinement of the partition of $k$-strong components of $G$.
\end{lemma}

\begin{definition}
    Given a weighted graph $G = (V, F, w)$ and an edge $f\in F$, the \emph{strength} of $f$, denoted by $k_f$, in $G$ is the maximum value of $k$ such that $f$ is contained in a $k$-strong component of $G$.
\end{definition}

Alternatively, the strength of an edge $f\in F$ is the largest minimum cut size among all induced subgraphs $G[X]$ that contain $f$, where $X$ ranges over all subsets of $V$. The following two claims give some properties of strength of edges in a graph.

\begin{claim} [Corollary 4.9 in \cite{BenczurK15}] \label{cla:dif_n-1}
    Given a weighted graph $G$ on $n$ vertices, there are at most $n-1$ distinct values of edge strengths.
\end{claim}

\begin{claim} [Lemma 4.11 in \cite{BenczurK15}] \label{cla:sum_n-1}
    For any weighted graph $G=(V,F,w)$ on $n$ vertices, $\sum_{f\in F}\frac{w(f)}{k_f}\leq n-1$.
\end{claim}

We can compute the strength of every edge in $G$  by computing the global min-cut of $(n-1)$ induced subgraphs of $G$~\cite{BenczurK15}. For the completeness of the argument, we prove the following lemma in Appendix~\ref{sec:strength}.

\begin{lemma} \label{lem:comp-strength}
    Given a weighted graph $G$ with $n$ vertices and $m$ edges. There is an algorithm that computes the strength of each edge in $\tilde{O}(mn)$ time with high probability.
\end{lemma}

The following cut counting lemma due to Karger~\cite{Karger:1993:GMR:313559.313605} gives an upper bound on the number of ``small cuts'' in a graph.
\begin{lemma}[Corollary 8.2 in~\cite{Karger:1993:GMR:313559.313605}]\label{lem:cut-counting}
    Given a weighted graph $G=(V,F,w)$ with minimum cut size $c$, for all integers $\alpha\geq 1$, the number of cuts of the graph with weight at most $\alpha c$ is at most $\card{V}^{2\alpha}$. We will refer to such cuts as {\em $\alpha$-cuts} throughout the paper.

\end{lemma}

\section{Construction of Near-linear Size Hypergraph Cut Sparsifiers}
\label{sec:overview}

Before describing our approach of creating hypergraph sparsifiers, we briefly review Benczúr and Karger's algorithm for graph sparsifiers \cite{benczur1996approximating,BenczurK15}. 

Given a graph $G=(V,F,w)$, they construct a sparsifier $\hat{G}$ as follows: for each edge $f \in F$, we include $f$ in $\hat{G}$ with probability $p_f = \tilde{O}(\frac{w(f)}{k_f})$ (i.e. its weight over its strength). Every edge $f$ that gets sampled is assigned a weight of $\hat{w}(f)=\frac{w(f)}{p_f}$ in $\hat{G}$ . By Claim~\ref{cla:sum_n-1}, the expected size of the sparsifier is $\tilde{O}(n)$. For any cut $C=(S,\bar{S})$ in the graph, the expected size of $\hat{w}(\delta_{\hat{G}}(S))$ is equal to $w(\delta_G(S))$. We need to give an upper bound of the probability that $\card{\hat{w}(\delta_{\hat{G}}(S)) - \expect{\hat{w}(\delta_{\hat{G}}(S))}} > \eps \expect{\hat{w}(\delta_{\hat{G}}(S))}$. By concentration bounds, the larger the size of $C$, the lower the probability that $\hat{w}(\delta_{\hat{G}}(S))$ is far from its expectation. By Lemma~\ref{lem:cut-counting}, if a graph has minimum cut size $c$, for any integer $\alpha$, the number of cuts of size at most $\alpha c$ is at most $n^{2\alpha}$. So we can group the cuts in different sizes based on this $\alpha$ value, take a union bound within each group, and then take a union bound over all groups to prove that with high probability, every cut in $\hat{G}$ has size close to its expectation. This gives a $(1 \pm \eps)$-approximate cut sparsifier.

Recently, Kogan and Krauthgamer~\cite{kogan2015sketching} generalized this approach to hypergraphs by defining an analogue of edge strengths for hyperedges. Most of the analysis for standard graphs also holds in the case of hypergraphs. The main difference is that in hypergraphs, the cut counting bound (Lemma~\ref{lem:cut-counting}) is no longer true. Instead, the authors prove that if the minimum cut size of a hypergraph is $c$, the number of cuts with size at most $\alpha c$ is $O(2^{\alpha r}n^{2\alpha})$ for any integer $\alpha$, where $r$ is the maximum cardinality of the edges in the hypergraph (see the footnote on page $2$ for an example showing that an exponential dependence on $r$  is necessary even for constant $\alpha$). This increase in the number of $\alpha$-cuts in turn requires edges to be oversampled at a rate that is $O(r)$ times higher, giving a hypergraph sparsifier of size $\tilde{O}(nr)$.

\iffalse
The authors of~\cite{kogan2015sketching} also give an example showing that for any $\alpha$ and $r$, there exist hypergraphs that indeed contain $\Omega(2^{\alpha r})$ cuts of size at most $\alpha c$. This suggests that it might not be possible to get a sparsifier with $o(nr)$ edges using their approch.

\begin{example} [\cite{kogan2015sketching}]
    Consider a ``sunflower'' graph defined as follows: there are $m$ hyperedges of size $r$ and $n = m(r-1)+1$ vertices. All edges contains the vertex $v_0$. For any two edges, their intersection is $\{v_0\}$. The minimum cut size of the graph is $1$. For any $1 \le i \le m$, let $V_i$ be the set of vertices contained in the $i^{th}$ edge except $v_0$. For any $1 \le \alpha \le m$, if a cut only cuts through at most $\alpha$ sets among $V_1,V_2,\dots,V_m$, the cut size is $\alpha$. Since each set $V_i$ has $r-1$ vertices, there are $2^{r-1}$ ways of cutting the set $V_i$. So there are at least $(2^{r-1})^\alpha \cdot \binom{m}{\alpha} = \Omega(2^{\alpha r})$ cuts have size $\alpha$ for any $\alpha \le m/2$.
\end{example}
\fi

\subsection{Overview of Our Approach}
Similar to the previous works on graph/hypergraph sparsification, for each edge $e$ in the hypergraph $H$, we will assign a probability $p_e$ of sampling the edge in the sparsifier $\hat{H}$. If $e$ is sampled, we give it weight $\frac{w_e}{p_e}$ in the sparsifier. However, unlike~\cite{kogan2015sketching}, our probabilities are not decided by the strength of the edge $e$ in $H$. Instead, we derive these probabilities from edge strengths in an auxiliary standard graph $G$, where for each hyperedge $e$ in $H$, we create a clique over the vertices of $e$ in $G$ such that the total weight of these clique edges is $w_e$. The hyperedge sampling probability $p_e$ is derived from the strengths of the edges in the associated clique in $G$. 

To prove that the sparsifier $\hat{H}$ is valid, we compare $\hat{H}$ to the Benczúr-Karger sparsifier $\hat{G}$ of $G$. For any cut $C$, it is not hard to see that the total weight of $C$ in $H$ is at least as large as the size of $C$ in $G$. Consider the cut size in $\hat{H}$ as the sum of several random variables (each one representing an edge/hyperedge across the cut). By concentration bounds, the higher the probability mass of these random variables, the greater is the concentration of their sum, which means the variance of the size of $C$ in $\hat{H}$ is at most its variance in $\hat{G}$. So we can use the cut-counting bound for standard graphs on $\hat{G}$ to analyze the concentration of the hypergraph sparsifier $\hat{H}$.

The approach of analyzing the performance of a hypergraph sparsifier through an auxiliary standard graph is also used in~\cite{BansalST19}. The authors use it to build a spectral sparsifier of a hypergraph. For a hyperedge $e$ in $H$, like~\cite{BansalST19}, a natural way of assigning its weight is to distribute its weight uniformly among all corresponding edges in $G$. However, this may cause the strengths of these edges in $G$ to be very different. Two natural ways of assigning $p_e$ are to either let $p_e$ be decided by the maximum inverse strength of these edges or decided by the average inverse strength. We can prove that deriving probabilities from the maximum inverse strength gives us small variance in cut sizes, while deriving probabilities from the average inverse strength results in a small number of sampled edges. However, the first approach may cause the number of sampled edges to be too large and the second approach cannot guarantee that the variance of the cut sizes in $\hat{H}$ is small enough. The two examples below illustrate this.

% \begin{figure}
%   \centering
%   \subfigure[]{\includestandalone[mode=buildnew]{example1}
%   } 
%   \subfigure[]{\includestandalone[mode=buildnew]{example2}} 
%   \caption{Examples \ref{example:1} and}
% \end{figure}

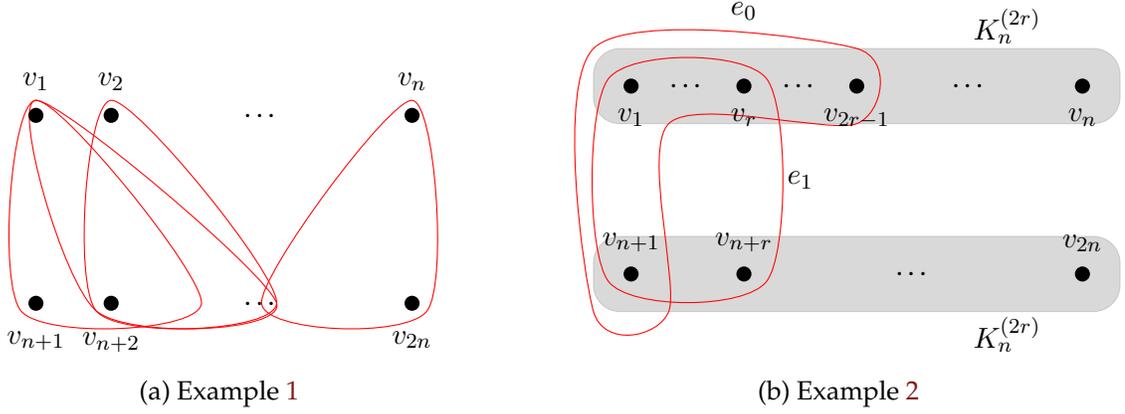
\begin{figure}[ht]
  \centering
  \begin{subfigure}[b]{0.5\linewidth}
    \centering\usetikzlibrary{topaths,calc}

\begin{tikzpicture}
    \node (v1) at (0,2.5) {};
    \node (v2) at (1,2.5) {};
    \node (v4) at (5,2.5) {};
    \node (v5) at (0,0) {};
    \node (v6) at (1,0) {};
    \node (v8) at (5,0) {};
    \path (v2) -- node[auto=false]{\ldots} (v4);
    \path (v6) -- node[auto=false]{\ldots} (v8);

    \draw [red] plot [smooth cycle] coordinates {($ (v1) + (0, 0.2)$) ($ (v5) + (-0.2, -0.1)$) ($ (v6) + (+1.2, -0.0)$)};
    \draw [red] plot [smooth cycle] coordinates {($ (v1) + (0, 0.2)$) ($ (v6) + (-0.2, -0.1)$) ($ (v6) + (+2.2, -0.0)$)};
    \draw [red] plot [smooth cycle] coordinates {($ (v2) + (0, 0.2)$) ($ (v6) + (-0.2, -0.1)$) ($ (v6) + (+2.2, -0.0)$)};
    %\draw [red] plot [smooth cycle] coordinates {($ (v2) + (0, 0.2)$) ($ (v5) + (-0.2+2, -0.0)$) ($ (v6) + (+3.2, -0.0)$)};

    \draw [red] plot [smooth cycle] coordinates {($ (v4) + (0, 0.2)$) ($ (v8) + (+0.2, -0.1)$) ($ (v8) + (-2, -0.0)$)};

    \fill (v1) circle (0.1) node [label={[xshift=0.0cm, yshift=0.07cm]$v_1$}] {};
    \fill (v2) circle (0.1) node [label={[xshift=0.0cm, yshift=0.07cm]$v_2$}] {};
    \fill (v4) circle (0.1) node [label={[xshift=0.0cm, yshift=0.07cm]$v_n$}] {};
    \fill (v5) circle (0.1) node [label={[label distance=0.1cm]270:$v_{n+1}$}] {};
    \fill (v6) circle (0.1) node [label={[label distance=0.12cm]270:$v_{n+2}$}] {};
    \fill (v8) circle (0.1) node [label={[label distance=0.1cm]270:$v_{2n}$}] {};

\end{tikzpicture}
    \caption{Example \ref{example:1}}
  \end{subfigure}%
  \begin{subfigure}[b]{0.5\linewidth}
    \centering\begin{tikzpicture}
    \node (v1) at (0,2.5) {};
    \node (v2) at (1.5,2.5) {};
    \node (v3) at (3,2.5) {};
    \node (v4) at (6,2.5) {};
    \node (v5) at (0,0) {};
    \node (v6) at (1.5,0) {};
    \node (v8) at (6,0) {};

    \fill [gray,rounded corners=10, draw, opacity=0.3, above] ($ (v1) + (-0.5, +0.5) $) --($ (v1) + (-0.5, -0.5) $) --($ (v4) + (+0.5, -0.5) $) --($ (v4) + (+0.5, +0.5) $) -- cycle;
    \node (top) at (5, 3.3) {$K_n^{(2r)}$};

    \fill [gray,rounded corners=10, draw, opacity=0.3, above] ($ (v5) + (-0.5, +0.5) $) --($ (v5) + (-0.5, -0.5) $) --($ (v8) + (+0.5, -0.5) $) --($ (v8) + (+0.5, +0.5) $) -- cycle;
    \node (top) at (5, -0.8) {$K_n^{(2r)}$};

    \draw [red] plot [smooth cycle] coordinates {($ (v5) + (-0.5, -0.5)$) ($ (v1) + (-0.5, 0.5)$) ($ (v3) + (0, 0.5)$) ($ (v3) + (0, -0.5)$) ($ (v1) + (0.5, -0.5)$) ($ (v5) + (+0.5, -0.5)$)};
    \node (e1) at ($ (v2) + (0, 1) $) {$e_0$};

    \draw [red] plot [smooth cycle] coordinates {($ (v5) + (-0.3, -0.1)$) ($ (v1) + (-0.3, 0.1)$) ($ (v2) + (0.3, 0.1)$) ($ (v6) + (0.3, -0.1)$)};
    \node (e1) at (2.25, 1.25) {$e_1$};

    % \draw [red] plot [smooth cycle] coordinates {($ (v1) + (0, 0.2)$) ($ (v5) + (-0.2, -0.1)$) ($ (v6) + (+1.2, -0.0)$)};
    % \draw [red] plot [smooth cycle] coordinates {($ (v1) + (0, 0.2)$) ($ (v6) + (-0.2, -0.1)$) ($ (v6) + (+2.2, -0.0)$)};
    % \draw [red] plot [smooth cycle] coordinates {($ (v2) + (0, 0.2)$) ($ (v6) + (-0.2, -0.1)$) ($ (v6) + (+2.2, -0.0)$)};
    % \draw [red] plot [smooth cycle] coordinates {($ (v2) + (0, 0.2)$) ($ (v5) + (-0.2+2, -0.0)$) ($ (v6) + (+3.2, -0.0)$)};

    % \draw [red] plot [smooth cycle] coordinates {($ (v4) + (0, 0.2)$) ($ (v8) + (+0.2, -0.1)$) ($ (v8) + (-2, -0.0)$)};

    \path (v1) -- node[auto=false]{\ldots} (v2);
    \path (v2) -- node[auto=false]{\ldots} (v3);
    \path (v3) -- node[auto=false]{\ldots} (v4);
    \path (v6) -- node[auto=false]{\ldots} (v8);
    \fill (v1) circle (0.1) node [label={[label distance=0.03cm]270:$v_1$}] {};
    \fill (v2) circle (0.1) node [label={[label distance=0.03cm]270:$v_r$}] {};
    \fill (v3) circle (0.1) node [label={[label distance=0.03cm]270:$v_{2r-1}$}] {};
    \fill (v4) circle (0.1) node [label={[label distance=0.03cm]270:$v_n$}] {};
    \fill (v5) circle (0.1) node [label={[label distance=0.03cm]90:$v_{n+1}$}] {};
    \fill (v6) circle (0.1) node [label={[label distance=0.03cm]90:$v_{n+r}$}] {};
    \fill (v8) circle (0.1) node [label={[label distance=0.03cm]90:$v_{2n}$}] {};

\end{tikzpicture}
    \caption{Example \ref{example:2}}
  \end{subfigure}
  \caption{Illustrations of Examples \ref{example:1} and \ref{example:2}. $K_n^{(2r)}$ refers to a copy of the complete $2r$-uniform hypergraph.}
\end{figure}

\begin{example}\label{example:1}
    Consider the following hypergraph with $2n$ vertices $v_1,v_2,\dots v_{2n}$: for any $1 \le i \le n$, we have all $\binom{n}{r-1}$ edges of size $r$ containing $v_i$ and $r-1$ vertices in $\{v_{n+1},v_{n+2},\dots,v_{2n}\}$. Suppose we were to distribute the weight of each hyperedge uniformly in the auxiliary graph $G$, each edge in $G$ has weight $1/\binom{r}{2} = O(1/r^2)$. For any $1 \le i \le n$, the weighted degree of $v_i$ in the graph $G$ is $O(1/r) \cdot \binom{n}{r-1}$, which means for each hyperedge, some of the edges in the associated clique in $G$ have strength $O(1/r) \cdot \binom{n}{r-1}$. Hence if the hyperedges are sampled according to the minimum strength of the corresponding edges in $G$, each hyperedge will be sampled with probability $\frac{\Omega(r)}{\binom{n}{r-1}}$, and the expected number of edges in the sparsifier will be $\Omega(nr)$ since there are $n \cdot \binom{n}{r-1}$ hyperedges.
\end{example}

\begin{example}\label{example:2}
    Consider the following hypergraph with $2n$ vertices and hyperedge size $2r \le \frac{n}{2}$: let $V = V_1 \cup V_2$ where $V_1 = \{v_1,\dots,v_n\}$ and $V_2 = \{v_{n+1},\dots,v_{2n}\}$. The graph contains one hyperedge $e_0 = \{v_1,\dots, v_{2r-1},v_{n+1}\}$, and one hyperedge $e_1 = \{v_1,\dots,v_r,v_{n+1},\dots,v_{n+r}\}$. There are also $\binom{n}{2r}$ hyperedges in $V_1$ and $\binom{n}{2r}$ hyperedges in $V_2$. Suppose we distribute the weight of each hyperedge uniformly in the auxiliary graph $G$. The cut size of $C=(V_1,V_2)$ is $\Theta(1)$ in $G$ since there are $r^2+2r-1$ edges of weight $1/\binom{2r}{2}$ crossing $C$. On the other hand, the induced subgraphs $G[V_1]$ and $G[V_2]$ both has minimum cut size $\Omega(2^r)$. So for any edge in $G$ crossing the cut $C$, its strength is $\Theta(1)$, and other edges in $G$ have strength $\Omega(2^r)$. Let $F_0$ be set of edges in $G$ corresponding to $e_0$. About $1/r$ fraction of the edges in $F_0$ have strength $\Theta(1)$ while the others have strength $\Omega(2^r)$. Both $\binom{r}{2} / (\sum_{f \in F_0} k_f)$ (inverse of average) and $(\sum_{f \in F_0} \frac{1}{k_f})/\binom{r}{2}$ (average of inverse) are $O(1/r)$. However, the cut $C$ has size $2$ in the hypergraph, which means that in order to build a $(1 \pm \eps)$-approximate cut sparsifier with $\eps < 1/2$, the edge $e_0$ must be included.
\end{example}

To solve this problem, we give an algorithm that assigns the weights of edges in $G$ such that for each hyperedge $e$, the strength of all corresponding edges in $G$ whose weight is positive is close to the smallest strength edge in the clique (we will formally define this idea in the next sub-section). In this case, the maximum inverse strength is quite close to the average inverse strength, so if $p_e$ is decided by the smallest strength (i.e. the largest inverse strength) in the clique, both the size of the sparsifier and the variance of the cuts have the properties we desire. 

\subsection{Construction of the Cut Sparsifier}
In this section, we formalize the ideas introduced in the previous section. To simplify the analysis, we first consider {\em unweighted }hypergraphs, and then give a simple reduction from the weighted case to the unweighted case. Later, in Section~\ref{sec:group}, we present a more sophisticated approach for handling weighted hypergraphs that gives us our final algorithm whose run-time has only a linear dependence on $m$.

Let $H = (V, E)$ be an unweighted multi-hypergraph with $|V|=n$ and $|E|=m$. Our goal is to create a $(1 \pm \eps)$-approximate cut sparsifier, given any $\eps \in (0,1]$. That is, we want to create a weighted hypergraph $\hat{H} = (V, \hat{E}, \hat{w})$ where $\hat{E} \subseteq E$ such that with high probability, for all cuts $C=(S,\bar{S})$ of $V$,
$$
\card{\hat{w}(\delta_{\hat{H}}(S))-\card{\delta_H(S)}} \leq \eps \card{\delta_H(S)}.
$$
In other words, the graph $\hat{H}$ preserves all cuts up to a factor of $(1 \pm \eps)$. We will sample the graph $\hat{H}$ by computing a probability $p_e$ for each edge $e\in E$. Each edge $e\in E$ is included in $\hat{H}$ with probability $p_e$, and if included, it is given a weight of $\hat{w}(e):=1/p_e$.

Given a hyperedge $e\in E$, define $F_e := \{\{u,v\}:u,v\in e, u\neq v\}$ as the clique on the vertex set of $e$. Let $F:=\bigcup_{e\in E}F_e$ be the \emph{multiset union} of all such cliques. Given a weight function $w^F:F\rightarrow\mathbb{R}_{\geq 0}$, we define $G=(V, F, w^F)$ as the weighted multigraph induced by $w^F$. Finally, given any subset $F_{sub}\subseteq F$, define $F_{sub}^+ = \{f\in F_{sub}:w^F(f)>0\}$ to be subset of $F_{sub}$ containing only positive weight edges.

For all hyperedges $e\in E$, define $\kappa_e := \min_{f\in F_e}k_f$ to be the minimum strength over \textit{all} edges in its associated clique, and $\kappa_e^{\max}:= \max_{f\in F_e^+}k_f$ to be the maximum strength over \textit{all positive-weighted} edges in its associated clique.

\begin{definition} \label{def:balance}
	Let $\gamma \geq 1$ be some parameter. The weight function $w^F:F\rightarrow \mathbb{R}_{\geq 0}$ is called a $\gamma$\emph{-balanced weight assignment} if it satisfies the following two conditions for all $e\in E$ in the hypergraph $H$:
	\begin{enumerate}
        \item[(1)] $\sum_{f\in F_e}w^F(f) = 1$, and
        \item[(2)] $\kappa_e^{\max}/\kappa_e\leq \gamma$.
    \end{enumerate}
\end{definition}

The next theorem, whose proof appears in Section~\ref{sec:balance}, shows that there exists a $\gamma$-balanced weight assignment for any $\gamma\ge 2$. We say two hyperedges are {\em distinct} if the vertex sets of these two hyperedges are not the same.

\begin{theorem}\label{thm:balance}
    Suppose we are given a hypergraph with $n$ vertices and $m$ hyperedges such that there are at most $\bar{m}$ distinct hyperedges. Then for any integer $\gamma \ge 2$, there is an algorithm that runs in $\tilde{O}(m\bar{m}n^4)$ time and finds a $\gamma$-balanced weight assignment.
\end{theorem}

In fact, with a more careful analysis, we can prove the statement of Theorem~\ref{thm:balance} is true for any real number $\gamma>1$. Together with Bolzano-Weierstrass theorem and some standard analysis, we can prove the existance of a balanced weight function even for $\gamma=1$. See Appendix~\ref{sec:perfect-balance} for more details.

Given such a weight assignment, the theorem below, whose proof appears in Section~\ref{sec:sparsifier}, shows that sampling with probabilities proportional to $1/\kappa_e$ gives a good sparsifier:

\begin{theorem}\label{thm:sparsifier}
    Let $\eps\in (0,1]$ and let $d$ be any integer constant. Suppose $w^F$ is a $\gamma$-balanced weight assignment of $H$. Consider a random subgraph $\hat{H}$ of $H$ where each edge $e\in E$ is sampled with probability $p_e:=\min(1,\frac{8(d+6)\gamma^2\log n}{0.38\eps^2\kappa_e})$
    and is given weight $1/p_e$ if sampled. Let $\hat{w}$ be this weight function on the sampled edges. Then with probability at least $1-O(n^{-d})$, for every cut $C=(S,\bar{S})$,
  $$
  \card{\hat{w}(\delta_{\hat{H}}(S)) - \card{\delta_H(S)}} \leq 2\eps \card{\delta_H(S)}.
  $$
  Furthermore, the expected number of edges in $\hat{H}$ is $O(\frac{\gamma^3 n\log n}{\eps^2})$.
\end{theorem}

Setting $\gamma=2$, for any unweighted hypergraph $H=(V,E)$, by Theorem~\ref{thm:balance}, there exists an algorithm that finds a $\gamma$-balanced weight assignment. Thus by Theorem~\ref{thm:sparsifier}, we can create a $(1 \pm \eps)$-approximate cut sparsifier of $H$ of size $O(\frac{n\log n}{\eps^2})$ with high probability.

%By Theorem~\ref{thm:balance} and Theorem~\ref{thm:sparsifier}, for any $\gamma\ge 2$, and an unweighted hypergraph $H=(V,E)$, there exists an algorithm that constructs a $(1 \pm \eps)$-approximate cut sparsifier of $H$ of size $O(\frac{\gamma n\log n}{\eps^2})$ in $m^{O(\log_{\gamma}m)}$ time with high probability. (We use Theorem~\ref{thm:balance} to find a $\gamma^{1/3}$-balanced weight assignment so that the final sparsifier has size $O(\gamma n\log n/\eps^2)$.) 

The corollary below gives a simple reduction from the weighted case to the unweighted case.

\begin{corollary} \label{cor:weighted}
    Given a weighted hypergraph $H=(V,E,w)$, suppose $W$ is the ratio of the largest edge weight to the smallest edge weight in $H$. Then for any $\eps\in (0,1]$, there exists an algorithm that constructs an $(1 \pm \eps)$-approximate sparsifier of $H$ with size $O(\frac{n\log n}{\eps^2})$ in $\tilde{O}(Wm^2n^4)$ time with high probability.
\end{corollary}

\begin{proof}
    Without loss of generality, assume that $1/\epsilon$ is an integer, and also that the weights $w$ are between $3/\epsilon$ and $3W/\epsilon$. For every edge $e\in E$, we add $\lfloor w(e)\rfloor$ copies of $e$ to a multiset $E'$. Since $w(e)\geq 3/\epsilon$, the number of copies of $e$ in $E'$ is $(w(e)\pm 1)$, which is within the range $(1\pm\epsilon/3)\cdot w(e)$. Let $\hat{H}$ be a $(1 \pm \epsilon/3)$-approximate cut sparsifier of $H'=(V,E')$ computed using Theorems \ref{thm:balance} and \ref{thm:sparsifier}. Then the weight of a cut in $\hat{H}$ is within a $(1 \pm \epsilon/3)^2$ factor (which is within the range $(1 \pm \epsilon)$) of its weight in $H$. In $H'$, there are at most $Wm$ hyperedges and there are at most $m$ hyperedges are distinct with each other. By Theorem~\ref{thm:balance}, the running time is $\tilde{O}(Wm^2n^4)$.
\end{proof}

We prove Theorem~\ref{thm:balance} in Section~\ref{sec:balance} and Theorem~\ref{thm:sparsifier} in Section~\ref{sec:sparsifier}. In Section~\ref{sec:group}, we speed up our algorithm so that the running time is linear in $m$ and eliminate the dependance of $W$, and thus prove Theorem~\ref{thm:main}.

\section{Finding a $\gamma$-balanced Assignment} \label{sec:balance}

In this section, we prove Theorem~\ref{thm:balance}, which shows that given an unweighted hypergraph $H=(V,E)$ with $n$ vertices and $m$ hyperedges, and for any integer $\gamma \ge 2$, we can find a $\gamma$-balanced assignment in polynomial time. Although we only consider the case when $\gamma$ is an integer for convenience, the argument can be easily generalized to the case when $\gamma$ is not an integer.

We find a $\gamma$-balanced assignment using an iterative algorithm. We start with the uniform weight assignment. In each step, say $e$ is an unbalanced hyperedge (i.e. $e$ violates condition (2) of Definition \ref{def:balance}) where $f_1$ and $f_2$ are the two edges in $F_e$ that ``witness'' $e$ being unbalanced, i.e. $f_1$ has positive weight and $k_{f_1} > \gamma k_{f_2}$. We move weight from $f_1$ to $f_2$. Informally (we will prove this later), the strength of $f_1$ can only decrease and the strength of $f_2$ can only increase as a result of this weight transition. There are two possible events that may happen if we keep moving weight from $f_1$ to $f_2$: either the strength of $f_1$ finally moves within a $\gamma$ factor of $f_2$; or we end up moving all the weight of $f_1$ to $f_2$, but $k_{f_1}$ is still larger than $\gamma k_{f_2}$. In either case, $f_1$ and $f_2$ are no longer a pair of ``witnesses'' to $e$ being unbalanced. We repeat this weight transfer until no unbalanced hyperedge remains.

Before we formally describe the algorithm, we first prove a lemma that shows how edge strengths in a graph change when we change the weight of an edge.

\begin{lemma} \label{lem:change}
    Let $G=(V, E, w)$ be a weighted graph, and let $G'=(V, E, w')$ be the weighted graph obtained from $G$ by increasing the weight of some edge $f$ by $\delta$. For any edge $f'$, denote by $k_{f'}$ and $k'_{f'}$ the strengths of $f'$ in $G$ and $G'$ respectively. Then for any edge $f'$,
    \begin{enumerate}
        \item $k_{f'}\le k'_{f'}\le k_{f'} + \delta$
        \item If $k'_{f'} > k_{f'}$, then $k_{f'}\ge k_f$ and $k'_{f'}\le k'_f$
    \end{enumerate}

    % % Given a weighted graph $G$ and an edge $f$ in $G$, suppose we increase (resp. decrease) the weight of $f$ by $\delta$. Then for any edge $f'$ in the graph, 
    % % \begin{enumerate}
    % % 	\item The strength of $f'$ does not decrease (resp. increase), and it increases (resp. decreases) by at most $\delta$.
    % % 	\item If the strength of $f'$ increases (resp. decreases) by a nonzero amount, the strength of $f'$ was at least (resp. at most) the strength of $f$ before the change, and the strength of $f'$ is at most (resp. at least) the strength of $f$ after the change.
    % \end{enumerate}
\end{lemma}

\begin{proof}
	% Let $G$ be the original graph, and let $G'$ be the modified graph. We first prove the lemma in the case that the weight of $f$ was increased. For any edge $\hat{f}$, we use $k_{\hat{f}}$ and $k'_{\hat{f}}$ to refer to the strength of $\hat{f}$ in $G$ and $G'$ respectively.

	%We now prove the first part of the lemma. 
    Let $f'$ be an edge, and let $G[X_{f'}]$ be the induced subgraph of $G$ that contains $f'$ and has minimum cut size $k_{f'}$. Since we only increase the weight of an edge $f$, the minimum cut size of $G'[X_{f'}]$ is at least $k_{f'}$, which means $k_{f'}\le k'_{f'}$. On the other hand, since the weight of $f$ is increased by $\delta$, the minimum cut size of any induced subgraph is increased by at most $\delta$. So $k'_{f'} \le k_{f'} + \delta$.
	
	Next, we prove the second part of the lemma. Let $f'$ be an edge, and suppose $k'_{f'} > k_{f'}$. Let $G'[X'_{f'}]$ be the induced subgraph of $G'$ that contains $f'$ and has minimum cut size $k'_{f'}$. Since $k'_{f'} > k_{f'}$, the minimum cut size of $G[X'_{f'}]$ is strictly less than $k'_{f'}$, which means $f$ is a part of some minimum cut of $G[X'_{f'}]$. In particular, this implies that $f$ is in $X'_{f'}$, so $k'_f$ is at least the minimum cut size of $G'[X'_{f'}]$, which is $k'_{f'}$.

    On the other hand, let $G[X_f]$ be the induced subgraph of $G$ that contains $f$ and has minimum cut size $k_f$. Consider the subgraph $G[X'_{f'}\cup X_f]$. Let $C=(S,\bar{S})$ be a minimum cut of this induced subgraph, and let $c$ be the size of $C$. Since this subgraph contains $f'$, by definition of strength, $c$ is at most $k_{f'}$. Note that $X'_{f'}$ and $X_f$ have nonempty intersection (they both contain the edge $f$). Therefore any cut of $X'_{f'}\cup X_f$ must either cut through $X_f$, or cut through $X'_{f'}$ but not $X_f$. In the case that $C$ cuts through $X'_{f'}$ but not $X_f$, $C$ does not cut through $f$, so it has size at most $c$ in $G'[X'_{f'}]$ (since the weight of all edges crossing $C$ stays the same). This implies that the minimum cut of $G'[X'_{f'}]$ is at most $c$, which means that $k'_{f'}\leq c\leq k_{f'}$, contradicting our assumption. So it must be the case that $C$ cuts through the vertex set $X_f$, which means $c$ is at least the minimum cut size of $G[X_f]$, and therefore $k_f\leq c\leq k_{f'}$.

    %On the other hand, let $G[X_f]$ be the induced subgraph of $G$ that contains $f$ and has minimum cut size $k_f$. Consider the subgraph $G[X'_{f'}\cup X_f]$. Let $C=(S,\bar{S})$ be a minimum cut of this induced subgraph, and let $c$ be the size of $C$. Since this subgraph contains $f'$, by definition of strength, $c$ is at most $k_{f'}$. If $C$ does not cut through the vertex set $X_f$, then $C$ does not cut through $f$, so it has size at most $c$ in $G'[X'_{f'}]$ (since the weight of edges crossing $C$ stays the same). This implies that the minimum cut of $G'[X'_{f'}]$ is at most $c$, which means that $k'_{f'}\leq c\leq k_{f'}$, contradicting our assumption. So $C$ does indeed cut through the vertex set $X_f$, which means $c$ is at least the minimum cut size of $G[X_f]$, and therefore $k_f\leq c\leq k_{f'}$.

    %For the case when we decrease the weight of $f$, we can reverse the role of old graph and new graph, and use the same argument as the proof of the case of increasing the weight of $f$.
\end{proof}

Our algorithm will maintain the invariant that all weights in the current weight assignment graph are integer multiples of some fixed $\delta > 0$, and the magnitude of each weight update will be exactly $\delta$. In such a graph, Lemma~\ref{lem:change} immediately implies that changing (increasing or decreasing) the weight of some edge $f$ by $\delta$ can only change the strength of an edge $f'$ if $f$ and $f'$ have the same strength both before and after the change.

\subsection{The Algorithm}

Now we describe the algorithm to find a $\gamma$-balanced assignment. Let $\delta = \frac{1}{n^2}$. First we assign the initial weights $w^{init}:F\rightarrow \mathbb{R}_{\geq 0}$ with the following constraint: the weight of each edge in $G$ is an integer multiple of $\delta$ and is at least $2\delta$. We can always do so because each hyperedge in $H$ has weight $1$, which is an integer multiple of $\delta$, and the number of edges in the clique associated with a hyperedge is at most $\binom{n}{2}$, which is less than $\frac{1}{2\delta}$. These initial weights give us a set of initial edge strengths $k^{init}_f$ of the weighted graph $G^{init} = (V, F, w^{init})$. Define $K_{0} := \min_{f\in F}k^{init}_f$, and define $\ell$ to be the smallest integer such that $K_0 \cdot \gamma^{\ell}$ is larger than $\max_{f\in F}k^{init}_f$. For each integer $0 \le i \le \ell$, define $K_i = K_0 \cdot \gamma^i$. Note that since the weights of all edges are integer multiples of $\delta$, the strength of each edge is also an integer multiple of $\delta$, which means $K_0$ is an integer multiple of $\delta$. Since $\gamma$ is an integer, all $K_i$ is also integer multiples of $\delta$. We partition the interval $I=[K_0, K_{\ell}]$ into subintervals $I_0,I_1, I_2, \ldots, I_{\ell}$, where $I_j := (K_{i-1},K_i]$ for $i>0$, and $I_0=\{K_0\}$. Note that $\max_{f\in F}k_f^{init}$ is at most the total weight of the edges and $K_{0}$ is at least $2\delta$, so $\ell$ is at most $\log_{\gamma}(n^2m) = O(\log m)$. We fix this partition for the rest of this section.

We use this partition $I_0, I_1, I_2, \ldots, I_{\ell}$ to determine how to iteratively modify these weights. Given a real number $x\in I$, we define $\ind(x)$ to be the integer $j$ such that $x\in I_j$. Given a weight function $w^F:F\rightarrow \mathbb{R}_{\geq 0}$ and the corresponding edge strengths $k:F\rightarrow \mathbb{R}_{\geq 0}$, we say that a hyperedge $e\in E$ is \emph{bad} in $G = (V, F, w^F)$ if there exist some $f,f'\in F_e$ such that $w^F(f')>0$ and $k_f < K_{\ind(k_{f'}) - 1}$. It is clear that if a hyperedge is not bad, then it is $\gamma$-balanced. We note that in general, as we update the weights, $k_f$ and $k_{f'}$ might not be contained in $I$ (so $\ind(k_f')$ might not be defined), but as it will turn out that we will maintain the invariant that all the edge strengths are always contained in $I$. We expand this definition to $\ind(e):=\ind(\max_{f \in F_e^+} k_f)$. Note that a hyperedge $e$ is bad if and only if $\kappa_e < K_{\ind(e)-1}$.

We run the following algorithm: while there exist bad hyperedges, we find a bad hyperedge $e$ with the maximum $\ind(e)$. Let $f,f' \in F_e$ be a pair that such $w^F(f')>0$ and $k_f < K_{\ind(k_{f'}) -1}$. We move $\delta$ weight from $f'$ to $f$.

\begin{algorithm} \label{alg-balance}
    \caption{An algorithm that eliminates all bad hyperedges}
\SetAlgoLined
    $w = w^{init}$\;
    \While{there exists some bad hyperedge} {
     Let $e$ be the one with maximum $\ind(e)$\;
     Let $f_{\min}:=\arg\min_{f\in F_e}k_f$ and $f_{\max}:=\arg\max_{f\in F_e^+}k_{f}$\;
     Let $k_{\min}$ and $k_{\max}$ to be the strengths of $f_{\min}$ and $f_{\max}$, respectively\; 
     Increase $w(f_{\min})$ by $\delta$ and decrease $w(f_{\max})$ by $\delta$\;
 }
 Return $w$\;
\end{algorithm}

Note that throughout the execution of the algorithm, the weight of each edge is an integer multiple of $\delta$, so the strength of each edge throughout the running of the algorithm is also an integer multiple of $\delta$. To prove the correctness of the algorithm, we first prove an important invariant that is maintained by the algorithm. 

\begin{claim}\label{cla:interval-contained}
	Let $i$ equal the value of $\ind(e)$ at some iteration of the while loop. For any edge $f$ whose strength increased	as a result of transferring the weights (Line 6), $\ind(k_f) < i$ after executing the transfer of weights. Also, no edge $f$ has strength less than $K_{0}$ after executing the transfer of weights.
\end{claim}

\begin{proof}
	Fix some iteration of the while loop, and let $i=\ind(e)$. By definition of $\ind(e)$ and $f_{\max}$, we have $\ind(k_{\max}) = \ind(e)$. On the other hand, since $e$ is a bad hyperedge, we have $k_{\min} < K_{i-1}$, which means $k_{\min} \le K_{i-1} - \delta$ since $k_{\min}$ is an integer multiple of $\delta$. By the first half of Lemma~\ref{lem:change}, $k_{f_{\min}}$ is increased by at most $\delta$, which implies that after the weight transfer, $k_{f_{\min}} \le K_{i-1}$. By the second half of Lemma~\ref{lem:change}, for any edge $f$ such that $k_f$ increases, $k_f \le k_{f_{\min}} \le K_{i-1}$, so $\ind(k_f) \le i-1$ after the weight transfer. This concludes the first part of the claim.

    Now we prove the second part of the claim inductively. Suppose that before we change the weights, no edge has strength less than $K_0$. Since $K_0\leq k_{\min} < K_{i-1}$, $i \ge 2$, so $k_{\max}\ge K_1 + \delta$. By the first half of Lemma~\ref{lem:change}, $k_{f_{\max}} \ge K_1$ after the weight transfer. By the second half of Lemma~\ref{lem:change}, for any edge $f$ such that $k_f$ decreases, $k_f \ge k_{f_{\max}} \ge K_1 > K_0$ after the weight change. So the second invariant still holds and this concludes the second part of the claim.
\end{proof}

Claim~\ref{cla:interval-contained} essentially proves that the interval $I=[K_0,K_{\ell}]$ (which was defined using the initial graph $G^{init}$) is the correct range of strengths to focus on. Algorithm 1 gives a $\gamma$-balanced assignment if it terminates since there would be no bad hyperedges. Therefore, to prove Theorem~\ref{thm:balance}, it is sufficient to prove that the running time of Algorithm 1 is $\tilde{O}(m\bar{m}n^4)$. We call the $t^{th}$ iteration of the while loop as iteration $t$. The following claim is another important invariant of Algorithm 1. 

\begin{claim} \label{cla:noninc}
    For any integer $i$, we define iteration $t_i$ as the earliest iteration that the bad hyperedge $e$ in the while loop has $\ind(e) \le i$. Then after iteration $t_i$, the total weight of edges that have strength larger than $K_{i-1}$ is non-increasing.
\end{claim}

\begin{proof}
    For any $t\ge 1$, we denote $e^t$ as the bad hyperedge in line 3 during iteration $t$. 
    We say a hyperedge $e'$ is \textit{very bad} if $\kappa_{e'} < K_{\ind(e')-1} - \delta$. We first prove that at any iteration starting from $t_i$, no hyperedge $e'$ with $\ind(e') > i$ is very bad. We prove it by contradiction. Suppose the statement is not true, and let $\bar{t}\geq t_i-1$ be the first iteration such that after iteration $\bar{t}$, a hyperedge $\bar{e}$ is very bad. At the beginning of iteration $t_i$, by the definition of $e^{t_i}$, no hyperedge $e'$ with $\ind(e') > i$ is bad, and hence no such hyperedge is very bad. So $\bar{t} \ge t_i$, which means at the beginning of iteration $\bar{t}$, no hyperedge $e'$ with $\ind(e') > i$ is very bad. There are two possible reasons that would cause $\bar{e}$ to become very bad: either $\ind(\bar{e})$ is increased or $\kappa_{\bar{e}}$ is decreased during the weight transfer in iteration $\bar{t}$.

    Suppose $\ind(\bar{e})$ increases during the weight transfer in iteration $\bar{t}$, and let $f \in F^{+}_{\bar{e}}$ be the edge that $\ind(f)$ increases. By Lemma~\ref{lem:change}, $k_f$ increases by at most $\delta$ during iteration $\bar{t}$. On the other hand, since $k_f$ is always an integer multiple of $\delta$, $k_f = K_{\ind(f)}$ at the beginning of iteration $\bar{t}$. Let $\hat{f} \in F_{e^{\bar{t}}}$ be the edge whose weight is increased during iteration $\bar{t}$. By Lemma~\ref{lem:change}, $k_{\hat{f}} = k_f = K_{\ind(k_f)}$ since $k_{\hat{f}}$ is an integer multiple of $\delta$. So at the beginning of iteration $\bar{t}$, $\ind(e^{\bar{t}}) \ge \ind(f) + 2$ since $e^{\bar{t}}$ is bad. This means 
    $$k_{\hat{f}} = K_{\ind(f)} < K_{\ind(f)+1} - \delta \le K_{\ind(e^{\bar{t}}) -1} -\delta$$ 
    where the first inequality is because $K_{\ind(f)+1} - K_{\ind(f)} \ge K_1 - K_0 = (\gamma -1)K_0 \ge 2\delta$. So $e^{\bar{t}}$ is very bad at the beginning of iteration $\bar{t}$, which contradicts the minimality of $\bar{t}$.

    Now consider the other possibility - $\kappa_{\bar{e}}$ decreases while $\ind(\bar{e})$ does not increase during weight transfer in iteration $\bar{t}$. By Lemma~\ref{lem:change}, $\kappa_{\bar{e}}$ is decreased by at most $\delta$ during iteration $\bar{t}$, which means that at the beginning of iteration $\bar{t}$, $\kappa_{\bar{e}} \le K_{\ind(\bar{e})-1} - \delta$. So $\bar{e}$ is a bad hyperedge. On the other hand, by Lemma~\ref{lem:change}, $\kappa^{\max}_{e^{\bar{t}}} = \kappa_{\bar{e}}$, so $\ind(e^{\bar{t}}) < \ind(\bar{e})$, which contradicts that $e^{\bar{t}}$ is the bad hyperedge which has the maximum index at the beginning of iteration $\bar{t}$.

    So at any time after iteration $t_i$, there is no very bad hyperedge $e'$ with $\ind(e') > i$.

    Since the algorithm only moves the weight from a high strength edge to a low strength edge, there is only one way that the total weight of the edges that has strength larger than $K_{i-1}$ increases: the strength of some edges increase from less than or equal to $K_{i-1}$ to larger than $K_{i-1}$. At the beginning of any iteration $t'$ after $t_i$, by Claim~\ref{cla:interval-contained}, if $\ind(e^{t'}) \le i$, any edge $f$ whose strength increases has $k_f \le K_{i-1}$. On the other hand, if $\ind(e^{t'}) > i$, $e^{t'}$ is not very bad, which means $\kappa_{e^{t'}} \ge K_i - \delta > K_{i-1}$. So any edge $f$ whose strength increases already has $k_f > K_{i-1}$ at the beginning of iteration $t'$. So the total weight of edges that has strength larger than $K_{i-1}$ is non-increasing.
\end{proof}

\begin{claim} \label{cla:iter}
    Algorithm 1 iterates in the while loop $\tilde{O}(mn^2)$ times.
\end{claim}

\begin{proof}
    Throughout the running of the algorithm, for any $1 \le j \le \ell-1$, we define a nonnegative potential function $W_j$ as follows: before iteration $t_j$, $W_j$ is always equal to $m$; after iteration $t_j$, $W_j$ equals the total weight of edges that have strength larger than $K_j$. Since the total weight of all edges is $m$, by Claim~\ref{cla:noninc}, all $W_j$'s are non-increasing throughout the running of the algorithm. On the other hand, for each iteration, suppose the bad hyperedge $e$ has $\ind(e) = i$. Note that this iteration cannot be before $t_i$. In this iteration, we transfer $\delta$ amount of weight from an edge whose strength is larger than $K_{i-1}$ to an edge whose strength is less than $K_{i-1}$. Furthermore, the edge whose weight increases does not have strength larger than $K_{i-1}$ after the weight change. So $W_{i-1}$ is decreased by at least $\delta$. Thus, in each iteration, no $W_j$ increases, and $W_i$ is decreased by at least $\delta$, which means there are at most $m * \ell / \delta = \tilde{O}(mn^2)$ iterations since $\ell = O(\log m)$.
\end{proof}

By Claim~\ref{cla:interval-contained} and Claim~\ref{cla:iter}, Algorithm 1 correctly outputs a $\gamma$-balanced weight assignment within a polynomial number of iterations.

\begin{proof}[Proof of Theorem~\ref{thm:balance}]
    The multi-graph $G$ contains $O(mn^2)$ edges, so computing the initial weight assignment takes $O(mn^2)$ time. 

    In each iteration of the while loop, we need to compute the strength of all edges in $G$ and find the bad hyperedges with maximum index. Note that if two edges share the same endpoints, their strengths are the same, so to compute the strength of the edges, we only need to compute the strength on a weighted complete graph $\bar{G}$ where for each pair of vertices $(u,v)$, the weight of edge $(u,v)$ is the sum of weights of edges whose endpoints are $u$ and $v$ in $G$. By Lemma~\ref{lem:comp-strength}, we need $\tilde{O}(n^3)$ time to compute the strength of all edges in $\bar{G}$ since there are $\binom{n}{2}$ edges in $\bar{G}$. Updating the weight of edges in $\bar{G}$ only takes $O(1)$ time. 

    Once the strengths of all edges in $\bar{G}$ has been computed, it takes $O(mn^2)$ time to check for each hyperedge if it is bad or not. However, if there are at most $\bar{m}$ distinct hyperedges, we can do it in $O(\bar{m}n^2)$ time in the following way: we group the hyperedges with the same vertex sets. For each group, we store the total weight in each edge slot, together with the identity of the hyperedges which have positive weight in each edge slot. To find a bad hyperedge with the maximum index in one group, we only need to consider the edge slot that has the maximum strength with positive weight, and check if the hyperedge that has weight in this slot is bad. In each iteration, it takes $O(\bar{m}n^2)$ time to find the maximum strength positive weight edge slot in each group and takes constant time to update the information in each edge slot.

    Thus overall, each iteration takes $\tilde{O}(\bar{m}n^2 + n^3) = \tilde{O}(\bar{m}n^2)$ time. So by Claim~\ref{cla:iter}, Algorithm 1 runs in $\tilde{O}(m\bar{m}n^4)$ time.
\end{proof}

\section{Constructing a Cut Sparsifier from a $\gamma$-balanced Assigment} \label{sec:sparsifier}

In this section, we prove Theorem~\ref{thm:sparsifier}, which shows that given a $\gamma$-balanced assigment $w^F$, we can construct a $(1 \pm \eps)$-approximate cut sparsifier that contains $O(\frac{\gamma^3n \log n}{\eps^2})$ edges. 

Let $\rho=\frac{8(d+6)\gamma^2\log n}{0.38 \eps^2}$, we sample each hyperedge $e$ in $H$ with probability $p_e = \min\{1,\frac{\rho}{\kappa_e}\}$. If an edge $e$ is sampled, it is assigned weight $\hat{w}_e = \frac{1}{p_e}$ in $\hat{H}$. We first show the expected number of edges in the sparsifier $\hat{H}$ is small.

\begin{claim}
    The expected number of edges in the sparsifier $\hat{H}$ is $O(\frac{\gamma^3 n\log n}{\eps^2})$.
\end{claim}
\begin{proof}
    The expected number of edges in the sparsifier is
    \begin{align*}
        \sum_{e\in E}p_e &\leq \rho\sum_{e\in E}\frac{1}{\kappa_e}=\rho \sum_{e\in E}\sum_{f\in F_e}\frac{w^F(f)}{\kappa_e} \\
                     &= \rho\sum_{e\in E}\sum_{f\in F_e}\frac{w^F(f)}{k_f}\frac{k_f}{\kappa_e} \leq \rho\gamma\sum_{e\in E}\sum_{f\in F_e}\frac{w^F(f)}{k_f} \\
                     &= \rho\gamma\sum_{f\in F}\frac{w^F(f)}{k_f}\leq \rho\gamma(n-1).
    \end{align*}

    For the second-to-last inequality, we used that for every $f\in F_e$ such that $w^F(f)>0$, $k_f\leq\kappa_e^{\max}\leq \gamma\kappa_e$ by Definition~\ref{def:balance}. The last inequality is due to Claim~\ref{cla:sum_n-1}, which asserts that $\sum_{f\in F}\frac{w^F(f)}{k_f}\leq n-1$. By the definition of $\rho$, this is $O(\gamma^3n\log n/\eps^2)$.
\end{proof}

In the rest of this section, we prove that $\hat{H}$ is indeed a good sparsifier. This proof is inspired by the framework of~\cite{benczur1996approximating}, who partition the edges into classes based on strength, and analyze the performance of each class separately. Before we start, as an additional piece of notation, given any subset of hyperedges $E'\subseteq E$, we define $\hat{E'}$ to be the subset of edges of $E'$ that were sampled in the sparsifier.

We first group the edges by their strengths. For each integer $i$, let $F_{\geq i} := \{ f\in F^+:k_f\geq \rho\cdot 2^i \}$ be the multiset of positive-weight edges with strength at least $\rho\cdot 2^i$. Let $E_{\geq i}:=\{e\in E:\kappa_e\geq \rho\cdot 2^i\}$ be the set of hyperedges with minimum strength at least $\rho\cdot 2^i$, and let $E_{\geq i}^{\max}:=\{e\in E:\kappa_e^{\max}\geq \rho\cdot 2^i\}$ be the set of hyperedges with maximum strength at least $\rho \cdot 2^i$. Note that $E_{\geq i}\subseteq E_{\geq i}^{\max}$.

Let $E_i:=E_{\geq i}\setminus E_{\geq i+1}$. We will prove an error bound for each $E_i$ separately. To prove this error bound, we define and analyze some slightly modified graphs. We first define some modified weights $w^F_i:F_{\geq i}\rightarrow \mathbb{R^+}$ and $w^E_i:E_{\geq i}\rightarrow \mathbb{R}^+$ in the following way: for an edge $f\in F$ such that $\rho\cdot 2^j\leq k_f < \rho\cdot 2^{j+1}$, $w^F_i(f):= w^F(f)\cdot 2^{i-j}$, and for a hyperedge $e\in E_j$, $w^E_i(e):=2^{i-j}$. Note that for a hyperedge $e\in E_i$, the weight of $e$ in $w^E_i$ remains $1$. Finally, define $G_{\geq i} = (V, F_{\geq i}, w_i^F)$, $H_{\geq i} = (V, E_{\geq i}, w_i^E)$, and $H_{\geq i}^{\max}=(V, E_{\geq i}^{\max}, w_i^E)$ to be the weighted graphs induced by these modified weights. 

The following lemma proves that for any $i$ and any cut $C$, the weight of the edges in $\hat{E}_i$ which cross $C$ is close to its expectation.

\begin{lemma}\label{cla:concentration}
    Fix some integer $i\geq 0$. With probability at least $1-4n^{-(d+1)}$, for all cuts $C=(S,\bar{S})$ of $V$, we have that
    $$
    \card{\hat{w}(\delta_{\hat{E}_i}(S)) - \card{\delta_{E_i}(S)}} \leq \frac{\eps}{\gamma}\cdot w^E_i(\delta_{E^{\max}_{\geq i}}(S))
    .$$
\end{lemma}

 Note that this lemma is not claiming that $\hat{E}_i$ is a good sparsifier of $E_i$ - the error term $\frac{\eps}{\gamma}w^E_i(\delta_{E^{\max}_{\geq i}}(S))$ can be much larger than $\eps\card{\delta_{E_i}(S)}$. We postpone the proof of Lemma~\ref{cla:concentration} and first show why Lemma~\ref{cla:concentration} completes the proof Theorem~\ref{thm:sparsifier}.

\begin{proof}[Proof of Theorem~\ref{thm:sparsifier}]
    In order to obtain concentration over all edges, we wish to take a union bound over every value of $i$ such that $E_i$ is not empty. By Claim~\ref{cla:dif_n-1}, there are at most $n-1$ such values of $i$.

     By Lemma~\ref{cla:concentration}, taking a union bound over these values of $i$, we get that with probability at least $1-4n^{-d}$, for all cuts $C=(S,\bar{S})$ of $V$ and for all $i$,
     \begin{align*}
         \card{\hat{w}(\delta_{\hat{E}_i}(S)) - \card{\delta_{E_i}(S)}} &\leq \frac{\eps}{\gamma}\cdot w_i^E(\delta_{E^{\max}_{\geq i}}(S)) \\
                                                                        &\leq \frac{\eps}{\gamma}\cdot\sum_{j\geq i-\log\gamma}2^{i-j}\card{\delta_{E_j}(S)} 
     \end{align*}
     where the last inequality is because $E_{\geq i}^{\max}\subseteq E_{\geq i-\log\gamma}$ (since $\kappa_e^{\max}/\kappa_e\leq \gamma$). Note that for all hyperedges $e$ that do not belong to any $E_i$, $\kappa_e\leq \rho$, so $p_e = 1$. That is, the contribution of these hyperedges to the error is $0$. We sum the errors over edges in $E_i$ for $i\geq 0$ to obtain that the total error is at most
     \begin{align*}
         & \sum_{i\geq 0}\card{\hat{w}(\delta_{\hat{E}_i}(S)) - \card{\delta_{E_i}(S)}} \\
         \leq & \frac{\eps}{\gamma}\sum_{i\geq 0}\sum_{j\geq i-\log\gamma}2^{i-j}\card{\delta_{E_j}(S)} \\
         = & \frac{\eps}{\gamma}\sum_{j\geq 0}\paren{\card{\delta_{E_j}(S)}\cdot\sum_{i\leq j+\log\gamma}2^{i-j}} \\
         \leq & 2\eps\sum_{j\geq 0}\card{\delta_{E_j}(S)},
     \end{align*}
     which is at most $2\eps \card{\delta_E(S)}$. Here the last inequality is due to $\sum_{i\leq j+\log\gamma}2^{i-j}\leq \sum_{i=-\floor{\log\gamma}}^{\infty} 2^{-i}\leq 2\gamma$. Therefore with probability at least $1-4n^{-d}$, for all cuts $C=(S,\bar{S})$, the size of $C$ in $\hat{H}$ is a $(1\pm 2\eps)$-approximation of the size in $H$.
\end{proof}

\subsection{Proof of Lemma~\ref{cla:concentration}}
Before proving Lemma~\ref{cla:concentration}, we first make some observations. As stated before, we associate the performance of $\hat{H}$ with the auxiliary standard graph $G$. The following claim states that for any cut $C$, the total weight of the edges crossing $C$ in $H_{\ge i}^{\max}$ is at least the total weights of the edges crossing $C$ in $G_{\ge i}$.

\begin{claim}\label{cla:cut-size}
    For any cut $C=(S,\bar{S})$ of $V$, $w_i^E(\delta_{E_{\geq i}^{\max}}(S)) \ge w_i^F(\delta_{F_{\geq i}}(S))$.
\end{claim}
\begin{proof}
    Let $e$ be some hyperedge, and let $f\in F_e$. If $f$ is a member of $G_{\geq i}$, then $e$ must be a member of $H_{\geq i}^{\max}$. Therefore if $f$ is cut by $C$ in $G_{\geq i}$, then $e$ must be cut by $C$ in $H_{\geq i}^{\max}$. Thus,
    \begin{align*}
        \sum_{f\in \delta_{G_{\geq i}}(S)}w_i^F(f) \leq \sum_{e\in \delta_{H_{\geq i}^{\max}}(S)}\sum_{f\in F_e}w_i^F(f) = \sum_{e\in \delta_{H_{\geq i}^{\max}}(S)} w_i^E(e).
    \end{align*}
    Here the equality is because for an edge $e\in E_j$, by condition (1) of $\gamma$-balanced weight assignments, $\sum_{f\in F_e}w_i^F(f) = \sum_{f\in F_e} w^F(f)\cdot 2^{i-j} = 2^{i-j} = w_i^E(e) $.
\end{proof}

In our analysis, we will independently bound the error incurred by each connected component of $G_{\ge i}$. The following claim states that no hyperedge is split among two different connected components of $G_{\ge i}$.

\begin{claim}\label{cla:same-component}
    For any $e\in E_{\geq i}$, the entire vertex set of $e$ belongs to the same connected component in $G_{\geq i}$.
\end{claim}
\begin{proof}
    Consider an edge $e\in E_{\geq i}$ and let $u,v$ be any two vertices in $e$. By definition of $\kappa_e$, the strength of the edge $(u,v)\in F_e$ is at least $\kappa_e$, so there exists some vertex set $X\subseteq V$ such that $u,v\in X$ and the induced subgraph $G[X]$ has min-cut size at least $\kappa_e>0$.

    Therefore $u$ and $v$ are connected by a path $P$ such that each edge on $P$ has positive weight. On the other hand, since $G[X]$ has min-cut size at least $\kappa_e$, which is at least $\rho \cdot 2^i$, all edges $f$ in $G[X]$ have $k_f\geq 2^i$. By definition of $F_{\geq i}$, this implies that all edges on $P$ are in $F_{\geq i}$, so $u$ and $v$ are connected in $G_{\geq i}$. 
\end{proof}

The following claim is similar to Lemma 3.2 in~\cite{benczur1996approximating}, which states that the min-cut size of each component in $G_{\ge i}$ is at least $\rho \cdot 2^i$, even with regards to the new weight function $w_i^F$. We give the proof of this claim for completeness.

\begin{claim}[Analog of Lemma 3.2 in \cite{benczur1996approximating}]\label{claim9.3.2}
    Let $A_G$ be a connected component of $G_{\geq i}$. Then the minimum cut size of $A_G$ is at least $\rho\cdot 2^i$.
\end{claim}
\begin{proof}
    Let $A'_G$ be the graph with the same vertex set and edge set as $A_G$, but instead of the modified weights $w^F_i$, we use the original weights $w^F$. We first claim that the strength of an edge $f$ in $A'_G$ is the same as its strength in $G$. To see this, let $X\subseteq V$ be a set of vertices such that $f\subseteq X$ and the induced weighted graph $G[X]$ has min-cut size at least $k_f$. Let $G[X]^+$ denote the subgraph of $G[X]$ that contains only positive-weight edges. Then every edge in $G[X]^+$ has strength at least $k_f\geq \rho\cdot 2^i$, which implies that $G[X]^+$ is a (induced) subgraph of $F_{\geq i}$. Since $G[X]^+$ is connected (and $A'_G$ is a connected component), $G[X]^+$ is also an induced subgraph of $A'_G$, providing a certificate that the strength of $f$ in $A'_G$ is $k_f$.

    Next, fix a cut $C=(S,\bar{S})$ of the vertex set of $A_G$. Let $f^*$ be a maximum strength edge in $\delta_{A'_G}(S)$. We claim that the total weight of strength $k_{f^*}$ edges in $\delta_{A'_G}(S)$ is at least $k_{f^*}$. To see this, let $X\subseteq V(A'_G)$ be a set of vertices such that $f^*\subseteq X$ and the min-cut size of $A'_G[X]$ is $k_{f^*}$. As required, all edges in $A'_G[X]$ have strength at least $k_{f^*}$, and the total weight of such edges crossing $C$ is at least $k_{f^*}$. Furthermore, by maximality of $f^*$, all edges crossing $C$ in $A'_G[X]$ have strength exactly $k_{f^*}$. Let $j$ be the index such that $\rho\cdot 2^j\leq k_{f^*}\leq \rho\cdot 2^{j+1}$. Now we bound the weight of edges crossing the cut in $A_G$:
    \begin{align*}
        w^F_i(\delta_{A_G}(S)) &\geq \sum_{f\in \delta_{A_G}(S):k_f = k_{f^*}} w^F_i(f) \\
                               &= \sum_{f\in \delta_{A_G}(S):k_f = k_{f^*}} w^F(f)\cdot2^{i-j} \\
                               &\geq k_{f^*}\cdot 2^{i-j}\geq \rho\cdot 2^i .
    \end{align*}
\end{proof}

To prove Lemma~\ref{cla:concentration}, for any cut $C=(S,\bar{S})$, we deal with each connected component in $G_{\ge i}$ separately. For each component $A_G$, we use concentration bound Lemma~\ref{chernoff} together with Claim~\ref{claim9.3.2} to prove that the total weights of the edges crossing $C$ in $A_G$ is preserved within an additive error $O(\max\{w_i^E(\delta_{A_H}(S)),w_i^F(\delta_{A_G}(S))\})$ where $A_H$ is the subhypergraph of $H_{\ge i}$ induced by the vertex set of $A_G$ (it is well defined due to Claim~\ref{cla:same-component}). On the other hand, since $w_i^E(\delta_{E_{\ge i}^{\max}}(S))$ dominates both $w_i^E(\delta_{E_{\ge i}}(S))$ and $w_i^F(\delta_{F_{\ge i}}(S))$ (by Claim~\ref{cla:cut-size}), by summing up the weights of the edges crossing $C$ in different components, we are able to prove that for the edges in $H_i$, the total weights of the edges crossing $C$ is perserved within additive error $O(w_i^E(\delta_{E_{\ge i}^{\max}}(S)))$. 

\begin{proof}[Proof of Lemma~\ref{cla:concentration}]
    Fix some connected component $A_G$ of $G_{\geq i}$, and let $V_A$ be the vertex set of this component. Let $C=(S,\bar{S})$ be some cut of $V_A$. For brevity, let $A_H:=H_{\geq i}[V_A]$ and $A_H':=H_i[V_A]$ be the subgraphs induced by this component.

    In order to apply Lemma \ref{chernoff}, we set the random variables $x_1,\ldots, x_k$ to be the sampled weights of edges in $\delta_{A_H'}(S)$ (so $k$ equals $\card{\delta_{A'_H}(S)}$). We set $N:=\max\{w^E_i(\delta_{A_H}(S)), w^F_i(\delta_{A_G}(S))\}$. We know that for each edge $e\in A_H'$, $w^E_i(e) = 2^{i-i}=1$, so $N\geq w^E_i(\delta_{A_H}(S))\geq \card{\delta_{A'_H}(S)}$. Therefore $N\geq k$, and we can indeed apply Lemma \ref{chernoff}.

    Let $c$ be the size of the minimum cut of $A_G$. By Claim \ref{claim9.3.2}, we have $c\geq \rho\cdot2^i$. Now define $\alpha:= \frac{w^F_i(\delta_{A_G}(S))}{c}$. Note that $N$ is at least $w^F_i(\delta_{A_G}(S))=\alpha c\geq \alpha \cdot \rho\cdot 2^i$. 

    Also, we have $\min_{e\in \delta_{A'_H}(S)}p_e = \min\{1, \min_{e\in \delta_{A'_H}(S)}\rho/\kappa_e\}\leq \min\{1, \rho/(\rho\cdot 2^{i+1})\}\leq 1/2^{i+1}$. The second-to-last inequality is because for any edge $e\in E_i$, we have that $\kappa_e\leq \rho\cdot 2^{i+1}$, and the last inequality is because $i\geq 0$.

    We apply Lemma \ref{chernoff} and get that
    \begin{align}\label{concentration-equation}
        \begin{split}
            &\prob{\card{\hat{w}(\delta_{\hat{A}_H'}(S)) - \card{\delta_{A'_H}(S)}} \geq \frac{\eps}{2\gamma} N} \\
            \leq & 2\exp{(-\frac{0.38\eps^2}{4\gamma^2}\cdot \min p_e\cdot N)} \\
            \leq & 2\exp{(-\frac{0.38\eps^2}{4\gamma^2}\cdot \frac{1}{2^{i+1}}\cdot \alpha\cdot\frac{8(d+6)\gamma^2\log n}{0.38\eps^2}\cdot 2^i)} \\
            = & 2n^{-(d+6)\alpha}.
        \end{split}
    \end{align}
    We now have a concentration bound which gets stronger as $\alpha$ increases.

    Apply cut counting bound (Lemma~\ref{lem:cut-counting}) on the weighted graph $A_G$, and we use this to apply a union bound over all cuts $C=(S,\bar{S})$ of $A_H$ such that $\alpha c\leq w_F^i(\delta_{A_G}(S))\leq 2\alpha c$ to conclude that with probability at least $1-2n^{2 \cdot 2\alpha}\cdot n^{-(d+6)\alpha} = 1-2n^{-(d+2)\alpha}$, the event in equation (\ref{concentration-equation}) does not occur for all of these cuts.
    We again apply the union bound over all values of $\alpha\geq 1$ that are powers of $2$ to obtain that with probability at least $1 - \sum_{j=0}^\infty 2n^{-(d+2)\cdot 2^j}\geq 1-4n^{-(d+2)}$, for all cuts $C=(S,\bar{S})$ of $V(A_H)$,
    \begin{align*}
        &\card{\hat{w}(\delta_{\hat{A}'_H}(S))- \card{\delta_{A'_H}(S)}} \\
        \leq & \frac{\eps}{2\gamma}\cdot \max\{w_i^E(\delta_{A_H}(S)),w_i^F(\delta_{A_G}(S))\} \\ 
        \leq & \frac{\eps}{2\gamma}\cdot \left(w_i^E(\delta_{A_H}(S))+w_i^F(\delta_{A_G}(S))\right).
    \end{align*}

    We now apply another union bound over all connected components of $G_{\geq i}$ (of which there are at most $n$) and sum this error term over all components. Let $C=(S,\bar{S})$ be a cut of the entire vertex set $V$. By Claim \ref{cla:same-component}, every hyperedge in $\delta_{H_{\geq i}}(S)$ is cut in exactly one such connected component. Therefore with probability at least $1-4n^{-(d+1)}$, for all cuts $C=(S,\bar{S})$ of $V$,
    \begin{align*}
        \card{\hat{w}(\delta_{\hat{E}_i}(S))- \card{\delta_{E_i}(S)}} \leq \frac{\eps}{2\gamma} \cdot \left(w_i^E(\delta_{E_{\geq i}}(S))+w_i^F(\delta_{F_{\geq i}}(S))\right).
    \end{align*}

    By Claim \ref{cla:cut-size} and by the fact that $H_{\geq i}$ is a subgraph of $H_{\geq i}^{\max}$, this is at most 
    $$
     \frac{\eps}{2\gamma}\cdot \left(2\cdot w_i^E(\delta_{E^{\max}_{\geq i}}(S))\right) = \frac{\eps}{\gamma}w_i^E(\delta_{E_{\geq i}^{\max}}(S)).
    $$
\end{proof}

\section{Speeding Up the Sparsifier Construction} \label{sec:group}

In this section, we complete the proof of Theorem~\ref{thm:main} by speeding up our algorithm so that its running time reduces to $\tilde{O}(mn+n^{10}/\eps^7)$ from $\tilde{O}(Wm^2n^4)$ (Corollary~\ref{cor:weighted}). Note that even for unweighted case ($W=1$), this is a significant speed-up in dense hypergraphs. 

At a high-level, the idea underlying the speed up is to reduce the general weighted problem to one where both $m$ and $W$ are  polynomially bounded in $n$. The first task is easy to accomplish using previously known results while the second task requires some additional ideas. 

Our starting point for reducing the number of edges is the following result by Chekuri and Xu~\cite{Chekuri018} which shows that the number of edges $m$ can be reduced to a polynomial in $n$ in near-linear time:

\begin{lemma} [Corollary 6.3 of~\cite{Chekuri018}] \label{lem:pre}
    A $(1 \pm \eps)$-approximate cut sparsifier of a weighted hypergraph $H$ with $O(n^3/\eps^2)$ edges can be found in $O(mn\log^2 n \log m)$ time with high probability.
\end{lemma}

After running this algorithm, we obtain a $(1 \pm \eps)$-approximate cut sparsifier of $H$ with only $O(n^3/\eps^2)$ edges.We then run the algorithm by Kogan and Krauthgamer~\cite{kogan2015sketching} and get a cut-sparsifier with $\tilde{O}(n^2/\eps^2)$ edges. 

\begin{lemma} [\cite{kogan2015sketching}] \label{lem:pre2}
    A $(1 \pm \eps)$-approximate cut sparsifier of a weighted hypergraph $H$ with $\tilde{O}(n^2/\eps^2)$ edges can be found in $O(mn^2 + n^3)$ time with high probability.
\end{lemma}

Since the number of hyperedges in the sparsifier given by Lemma~\ref{lem:pre} is $O(n^3/\eps^2)$, we only need $\tilde{O}(n^5/\eps^2)$ time to run the algorithm in Lemma~\ref{lem:pre2}. Let $\bar{H} = (V,\bar{E},\bar{w})$ be the sparsifier. 

It is worth noting that although the number of edges in $\bar{H}$ is polynomial, the ratio of maximum and minimum weight is still unbounded. In fact, even if $H$ is unweighted, the ratio of maximum and minimum weight of $\bar{H}$ still could be as large as $2^n$. To solve this problem, we group the edges by their weights. Let $\alpha = \frac{10n^2}{\eps^3}$ and $\bar{E} = E_1 \cup E_2 \cup \dots$ where $E_i=\{e \in \bar{E} : \bar{w}(e) \in [w_0 \cdot \alpha^{i-1},w_0 \cdot \alpha^i)\}$ where $w_0$ is the minimum weight in $\bar{H}$. 

Let $H_i = (V,E_i,\bar{w})$ and $m_i = \card{E_i}$. By Corollary~\ref{cor:weighted}, we only need $\tilde{O}(\alpha m_i^2 n^4)$ time to build a near-linear size (in $n$) sparsifier for each of $H_i$. However, if we combine these sparsifiers together, the size is no longer near-linear.

Note that $\alpha \ge \frac{10\bar{m}}{\eps}$ where $\bar{m}$ is the number of edges in $\bar{H}$. Suppose a cut separates an edge $e$ in $H_i$, the sum of weights of all edges in $\cup_{j \le i-2} E_j$ is less than $\eps/10$ fraction of the size of the cut. Therefore, for any $i$, we can ignore the performance of the sparsifier of $H_j$ for $j \le i-2$ within the connected components of $H_i$.

Define $E_{odd} = E_1\cup E_3\cup \ldots$, and $E_{even} = E_2\cup E_4\cup\ldots$. We will independently construct sparsifiers of $H_{odd} = (V, E_{odd}, w)$ and $H_{even} = (V, E_{even}, w)$ and merge them into a single sparsifier for $\bar{H}$.

\begin{lemma} \label{lem:faster-weighted}
    For any $0<\eps<1$, there is an algorithm that constructs $(1 \pm \eps)$-approximate cut sparsifiers for both $H_{even}$ and $H_{odd}$ with size $O(\frac{n\log n}{\eps^2})$ in $\tilde{O}(n^{10}/\eps^7)$ time with high probability.
\end{lemma}

Without loss of generality, we focus on $H_{even}$. The algorithm builds sparsifiers for each of $H_{2i}$ one by one from higher $i$ to lower $i$. Let $E_{> 2i} = \cup_{j>i} E_{2j}$ and $H_{>2i} = (V,E_{>2i},\bar{w})$. For each $i$, we first find all connected components of $H_{>2i}$. Let $V_{2i}^C$ be a vertex set such that each connected component (including isolated vertices) of $H_{>2i}$ is a ``supervertex'' in $V_{2i}^C$. Let $E^C_{2i}$ be the hyperedge set such that for each edge $e \in E_{2i}$, $E^C_{2i}$ contains the hyperedge $e'\subseteq V^C_i$ with weight $\bar{w}(e')=\bar{w}(e)$ that contains all vertices in $V^C_i$ such that $e$ contains a vertex in the corresponding connected component. Let $H_{2i}^C = (V^C_{2i},E^C_{2i},\bar{w})$.

For each connected component of $H_{2i}^C$, we build a $(1 \pm \frac{\eps}{2})$-approximate cut sparsifier by the algorithm in Corollary~\ref{cor:weighted}. We take the union of these sparsifiers and get an $\frac{\eps}{2}$-sparsifier $\hat{H}^C_{2i}=(V^C_{2i},\hat{E}^C_{2i},\hat{w})$ of $H^C_{2i}$. Let $\hat{H}_{2i} = (V,\hat{E}_{2i},\hat{w})$ be the graph ``restored'' from $\hat{H}^C_{2i}$, i.e. for each edge $e$ in $E_{2i}$, $e$ is in $\hat{E}_{2i}$ if the corresponding edge $e'$ is in $\hat{H}^C_{2i}$. It also gets the same weight as $e'$ if it is included in $\hat{H}_{2i}$. For any cut $(S,\bar{S})$ of $V_{2i}$ which does not cut any component in $H_{>2i}$, the cut size in $\hat{H}_{2i}$ and $\hat{H}^C_{2i}$ are the same, and the cut size in $H_{2i}$ and $H^C_{2i}$ are the same. In particular, this implies that $\hat{H}_{2i}$ is a good sparsifier of $H_{2i}$ with respect to all cuts that do not cut any component in $H_{>2i}$.

We output $\hat{H}_{even} = \cup_i \hat{H}_{2i}$ as a sparsifier of $H_{even}$. By Corollary~\ref{cor:weighted}, the running time is
\begin{align*}
    \sum_i \tilde{O}(\alpha m_i^2 n^4) = \tilde{O}((\sum_i m_i)^2\alpha n^4) = \tilde{O}(\alpha \bar{m}^2n^4) = \tilde{O}(n^{10}/\eps^7).
\end{align*}

We now prove $\hat{H}_{even}$ is indeed a good cut sparsifier of $H_{even}$. From this point on, we assume the algorithm in Corollary~\ref{cor:weighted} is always successful throughout the algorithm (which happens with high probability). We first prove that $\hat{H}_{even}$ is indeed a $(1 \pm \eps)$-approximate cut sparsifier of $H_{even}$.

\begin{claim} \label{cla:group-eps}
    $\hat{H}_{even}$ is a $(1 \pm \eps)$-approximate cut sparsifier of $H_{even}$.
\end{claim}

\begin{proof}
    We first prove that for any $i$, $\hat{w}(\hat{E}_{2i}) \le 3 \bar{w}(E_{2i})$. Equivalently, we prove that $\hat{w}(\hat{E}^C_{2i}) \le 3\bar{w}(E^C_{2i})$. Let $(S', \bar{S}')$ be some cut of $\hat{H}_{2i}^C$ of weight at least $\hat{w}(\hat{E}^C_{2i})/2$. Such a cut must exist because the expected weight of a random cut of a graph/hypergraph is at least half of the total weight of the graph. Since $\hat{H}_{2i}^C$ is a $(1 \pm \frac{\eps}{2})$-approximate cut sparsifier of $H_{2i}^C$, $\hat{w}(\delta_{\hat{H}_{2i}^C}(S'))\leq (1+\frac{\eps}{2})\cdot \bar{w}(\delta_{H_{2i}^C}(S'))\leq 1.5 \cdot \bar{w}(E_{2i}^C)$ since $\eps<1$. Therefore $\hat{w}(\hat{E}_{2i}^C)/2 \leq 1.5 \cdot \bar{w}(E_{2i}^C)$, concluding the proof.

    Now fix any cut $C=(S,\bar{S})$ of $V$. Let $i$ be the largest integer such that $\delta_{E_{2i}}(S) \neq \emptyset$. Since $\alpha \ge \frac{10\bar{m}}{\eps}$, $\bar{w}(\delta_{E_{2i}}(S))$ is at least $(1 - \frac{\eps}{10})$ fraction of $\bar{w}(\delta_{E_{even}}(S))$.

    Since $C$ does not cut through any component of $H_{>2i}$, $\hat{w}(\delta_{\hat{H}_{2i}}(S))$ is within $(1 \pm \frac{\eps}{2})$ fraction of $\hat{w}(\delta_{H_{2i}}(S))$, which means 
    \begin{align*}
        \hat{w}(\delta_{\hat{H}_{even}}(S)) \ge \hat{w}(\delta_{\hat{H}_{2i}}(S)) \ge (1-0.5\eps)\bar{w}(\delta_{H_{2i}}(S)) \ge (1-\eps)\bar{w}(\delta_{\bar{H}_{even}}(S)).
    \end{align*}

    On the other hand, since $\alpha \ge \frac{10\bar{m}}{\eps}$ and $\hat{w}(\hat{E}_{2j}) \le 3 \bar{w}(E_{2j})$ for any $j$, we have $\hat{w}(\cup_{j<i} \hat{E}_{2j}) < 0.3 \eps \cdot \bar{w}(\delta_{H_{even}}(S))$. which means
    \begin{align*}
        \hat{w}(\delta_{\hat{H}_{even}}(S)) 
        &\le \hat{w}(\delta_{\hat{H}_{2i}}(S)) + 0.3\eps \cdot \bar{w}(\delta_{E_{even}}(S)) \\
        &\le (1+0.5\eps)\bar{w}(\delta_{H_{2i}}(S)) + 0.3\eps \cdot \bar{w}(\delta_{E_{even}}(S)) \\
        &\le (1+\eps)\bar{w}(\delta_{\bar{H}_{even}}(S)).
    \end{align*}
\end{proof}

The next claim shows that $\hat{H}_{even}$ has near linear size.

\begin{claim} \label{cla:group-size}
    The size of $\hat{H}_{even}$ is $O(\frac{n \log n}{\eps^2})$.
\end{claim}

\begin{proof}
    For any $i>0$, let $\Delta_i = \card{V_{>2i}} - \card{V_{>2(i-1)}}$ for all $i>0$ and let $\card{V_{>0}}$ be the number of connected components in $H_{even}$. To prove the claim, it is sufficient to prove that $\card{\hat{E}_{2i}} = O(\frac{\Delta_i\log n}{\eps^2})$ for all $i>0$. 

    Suppose there are $\ell$ connected components in $H^C_{2i}$ and their sizes are $n_{i1},n_{i2},\dots,n_{i\ell}$. For any $j$, if $n_{ij}>1$, then $2(n_{ij}-1) \ge n_{ij}$, so the size of the sparsifier of this component is $O(\frac{(n_{ij}-1) \log n}{\eps^2})$ by Corollary~\ref{cor:weighted}. On the other hand, if $n_{ij}=1$, the component is an isolated vertex and we do not need to find a sparsifier for this component. So the total size of these sparsifiers is $\card{\hat{E}_{2i}} = O(\frac{\sum_{j=1}^{\ell} (n_{ij}-1) \log n}{\eps^2})$.

    For each component of $H^C_{2i}$ of size $n_{ij}$, the vertices in the component will contract to one single vertex in $V^C_{>2(i-1)}$, which means
    $$
    \card{V^C_{>2(i-1)}} = \ell = \sum_{j=1}^{\ell} (n_{ij}-(n_{ij}-1)) = \card{V^C_{>2i}} - \sum_{j=1}^{\ell} (n_{ij}-1).
    $$
    So $\sum_{j-1}^{\ell}(n_{ij}-1) = \Delta_i$, implying that $\card{\hat{E}_{2i}} = O(\frac{\Delta_i\log n}{\epsilon^2})$.
\end{proof}

Lemma~\ref{lem:faster-weighted} immediately follows from Claim~\ref{cla:group-eps} and Claim~\ref{cla:group-size}. Now we are ready to prove Theorem~\ref{thm:main}.

\begin{proof}[Proof of Theorem~\ref{thm:main}]
    We first apply the algorithm in Lemma~\ref{lem:pre} and Lemma~\ref{lem:pre2} to build $\bar{H}$, which runs in time $\tilde{O}(mn+n^5/\eps^2)$. Then we build the graphs $H_{even}$ and $H_{odd}$, find $(1 \pm \eps)$-approximate cut sparsifiers with size $O(\frac{n \log n}{\eps^2})$ for each of them and take the union of these two sparsifiers to get a $(1 \pm \eps)$-approximate cut sparsifier $\hat{H}$ of $\bar{H}$. By Lemma~\ref{lem:faster-weighted}, this runs in time $\tilde{O}(n^{10}/\eps^7)$. So we get a $(1 \pm O(\eps))$-approximate cut sparsifier $\hat{H}$ of $H$ with size $O(\frac{n \log n}{\eps^2})$, in $\tilde{O}(mn+n^{10}/\eps^7)$ time.
\end{proof}

\section*{Acknowledgements}

We would like to thank Chandra Chekuri for valuable discussions in the early stages of this work. This work was supported in part by NSF awards CCF-1617851, CCF-1763514, CCF-1934876, and CCF-2008305.

\appendix
\section{Appendix}
\subsection{Proof of Existence of 1-balanced Assignment} \label{sec:perfect-balance}

In this section, we prove that there exists a 1-balanced weight assignment $G=(V,F,w)$ for every hypergraph $H=(V,E)$. To do this, we first prove that the conclusion of Theorem \ref{thm:balance} holds for all $\gamma>1$ (as opposed to $\gamma\geq 2$). Equivalently, we prove that Theorem \ref{thm:balance} holds for $\gamma = 1 + 1/i$ every positive integer $i$. The only change needed in the algorithm is to use $\delta = \frac{1}{n^2i}$ instead of $\delta = \frac{1}{n^2}$, and to ensure that $K_0$ is at least $2i\delta$ instead of $2\delta$. The rest of the proofs are completely analogous, with the only modification being that $(\gamma - 1)K_0\geq 2\delta$ no longer follows from the fact that $\gamma \geq 2$, but simply from the fact that $K_0$ is at least $2i\delta = \frac{2\delta}{\gamma - 1}$. Note that the number of iterations (and hence the running time) of the algorithm is increased by a factor of $i^2$, since $\delta$ and $\ell$ are decreased and increased by a factor of $i$ respectively.

For the rest of this section, it will be convenient to represent a weight assignment $w:F\rightarrow\mathbb{R}_{\geq 0}$ as a vector in $\mathbb{R}_{\geq 0}^{|F|}$. Additionally, given a vector $w\in \mathbb{R}_{\geq 0}^{|F|}$, we use $k_f(w)$, $\kappa_e(w)$, and $\kappa^{\max}_e(w)$, and $F_e^+(w)$ to denote the value of these quantities in the weight assignment represented by $w$.

Let $\{w_i\in \mathbb{R}_{\geq 0}^{|F|}\}$ be a sequence of vectors such that $w_i$ represents a $1+1/i$-balanced weight assignment. We invoke the Bolzano-Weierstrass Theorem on this sequence:

\begin{theorem}[Bolzano-Weierstrass Theorem] 
Every bounded sequence of vectors in $\mathbb{R}^n$ has a convergent subsequence.
\end{theorem}

Denote this convergent subsequence by $\{w_i'\in \mathbb{R}^{|F|}\}$, and let $w$ be the limit of this subsequence. We will use a limiting argument to show that $w$ is $1$-balanced. First we note that the strength of an edge $k_f$ is a (1-Lipschitz) continuous function of the weight assignment. This follows immediately from the first half of Lemma \ref{lem:change}. Therefore $\lim k_f(w_i') = k_f(\lim w_i') = k_f(w)$. Since $\min$ is also a continuous function, this implies that 
\begin{equation}\label{eqn:lim-kappa}
    \begin{split}
    &\lim \kappa_e(w_i') = \lim\min_{f\in F_e}k_f(w_i') = \min_{f\in F_e}\lim k_f(w_i') \\
    =& \min_{f\in F_e}k_f(w) = \kappa_e(w)
    \end{split}
\end{equation}

We would like to be able to make a similar statement about $\lim\kappa^{\max}_e(w_i')$, but it is not true in general because $\kappa^{\max}_e$ is not a continuous function of the weight assignment vector. Instead, we observe that for $i$ large enough, the set $F_e^+(w_i')$ is a superset of $F_e^+(w)$, since the weight of any edge in $F_e^+(w)$ must eventually become positive in the sequence $\{w_i'\}$. So \begin{align}\lim \kappa^{\max}_e(w_i') = \lim \max_{f\in F_e^+(w_i')}k_f(w_i')\geq \lim \max_{f\in F_e^+(w)}k_f(w_i')\nonumber\\\label{align:lim-kappa-max}
 = \max_{f\in F_e^+(w)}\lim k_f(w_i') = \max_{f\in F_e^+(w)}k_f(w) = \kappa^{\max}_e(w)\end{align}
 
Here the inequality used the fact that for for large $i$, $F_e^+(w'_i)\supseteq F_e^+(w)$, and second equality used that $\max$ is a continuous function. Combining Equations \ref{eqn:lim-kappa} and \ref{align:lim-kappa-max}, \[\kappa^{\max}_e(w)\leq \lim \kappa_e^{\max}(w_i')\leq \lim ((1+1/i)\cdot \kappa_e(w_i')) = 1\cdot \kappa_e(w),\]
 where the second inequality holds because $w_i'$ is $1+1/i$-balanced. Therefore, $w$ is $1$-balanced, as desired.

\subsection{Computing Exact Edge Strengths} \label{sec:strength}

In this section, we give for completeness an algorithm that computes the {\em exact} strength of each edge in a graph and prove Lemma~\ref{lem:comp-strength}. Our algorithm will use as a subroutine
the following global min-cut result of Karger~\cite{Karger00}:

\begin{theorem} [\cite{Karger00}] \label{thm:glo-min-cut}
    Given a weighted graph $G$ with $n$ vertices and $m$ edges, there is a randomized algorithm that finds the minimum cut in $\tilde{O}(m)$ time with high probability.
\end{theorem}

The algorithm for computing exact edge strengths works as follows. 
We start by finding a minimum cut in the input graph $G$, and removing the edges in the minimum cut.
We then repeat this process in each connected component, until the graph becomes an empty graph. Now for each edge in the graph, we output the strength of this edge as the largest min-cut value among all connected components containing this edge, that are encountered during the execution of the algorithm.
 
\begin{lemma*}[Restatement of Lemma~\ref{lem:comp-strength}]
    Given a weighted graph $G$ with $n$ vertices and $m$ edges, there is a randomized algorithm that computes the strength of each edge exactly in $\tilde{O}(mn)$ time with high probability.
\end{lemma*}

\begin{proof}
    The above algorithm requires $(n-1)$ computations of global min-cut. Thus by Theorem~\ref{thm:glo-min-cut}, the total running time is $\tilde{O}(mn)$. We now prove that it correctly outputs exact edge strengths. We fix an edge $e$, let $\bar{k}_e$ denote the strength that our algorithm outputs for the edge $e$. It is clear that $\bar{k}_e \le k_e$ since by the definition of $\bar{k}_e$, there is a subgraph of $G$ which contains $e$ and has min-cut size $\bar{k}_e$. To show that $\bar{k}_e \ge k_e$ also holds, consider the induced subgraph $G[X]$ which contains the edge $e$ and has min-cut $k_e$. 
    During the execution of our algorithm, let $G[\bar{X}]$ be the last connected component encountered which fully contains $X$. By our choice of $G[\bar{X}]$, the min-cut of $G[\bar{X}]$ must also cut through $G[X]$, which means that the cut size in this step is at least the min-cut size of $G[X]$, which is $k_e$. Thus by the definition of $\bar{k}_e$, the value of $\bar{k}_e$ is at least the size of min-cut of $G[\bar{X}]$ since $G[\bar{X}]$ contains $G[X]$ which contains $e$. So $\bar{k}_e \ge k_e$.
\end{proof}

\bibliographystyle{abbrv}
\bibliography{general}

\end{document}